\documentclass[fleqn,usenatbib]{mnras}

\usepackage{newtxtext,newtxmath}
\usepackage{array}  

\usepackage[T1]{fontenc}

\DeclareRobustCommand{\VAN}[3]{#2}
\let\VANthebibliography\thebibliography
\def\thebibliography{\DeclareRobustCommand{\VAN}[3]{##3}\VANthebibliography}

\usepackage{graphicx}	
\usepackage{amsmath}	
\usepackage{makecell}
\usepackage{xcolor}

\newcommand{\Msol}{\textup{M}_\mathrm{\sun}}

\newcommand{\Mstar}{M_{\star}}

\title[Dark matter core creation and star formation]{EDGE: Dark matter core creation depends on the timing of star formation}

\author[C. Muni et al.]{
Claudia Muni$^{1}$\thanks{E-mail: claudia.muni.21@ucl.ac.uk},
Andrew Pontzen$^{2}$, 
Justin I. Read$^{3}$, 
Oscar Agertz$^{4}$, 
Martin P. Rey$^{5,6}$,
Ethan Taylor$^{3}$, 
\newauthor 
Stacy Y. Kim$^{7}$, 
and Emily I. Gray$^{3}$
\\
$^{1}$Department of Physics $\&$ Astronomy, University College London, Gower Street, London WC1E 6BT, UK\\
$^{2}$Institute for Computational Cosmology, Department of Physics, Durham University, South Road, Durham, DH1 3LE, UK\\
$^{3}$Department of Physics, University of Surrey, Guildford, GU2 7XH, Surrey, UK\\
$^{4}$Lund Observatory, Division of Astrophysics, Department of Physics, Lund University, Box 43, SE-221 00 Lund, Sweden\\
$^{5}$ Sub-department of Astrophysics, University of Oxford, DWB, Keble Road, Oxford OX1 3RH, UK \\ 
$^{6}$ University of Bath, Department of Physics, Claverton Down, Bath, BA2 7AY, UK \\
$^{7}$Carnegie Theoretical Astrophysics Center, Carnegie Observatories, 813 Santa Barbara St, Pasadena, CA 91106, USA
}

\date{Accepted XXX. Received YYY; in original form ZZZ}

\pubyear{2024}

\begin{document}
\label{firstpage}
\pagerange{\pageref{firstpage}--\pageref{lastpage}}
\maketitle

\begin{abstract}
We study feedback-driven cold dark matter core creation in the EDGE suite of radiation-hydrodynamical dwarf galaxy simulations. Understanding this process is crucial when using observed dwarf galaxies to constrain the particle nature of dark matter. While previous studies have shown the stellar-mass to halo-mass ratio $(\textrm{M}_{\star} / \textrm{M}_\mathrm{200})$ determines the extent of core creation, we find that in low-mass dwarfs there is a crucial additional effect, namely the timing of star formation relative to reionisation. Sustained post-reionisation star formation decreases central dark matter density through potential fluctuations; conversely, pre-reionisation star formation is too short-lived to have such an effect. In fact, large stellar masses accrued prior to reionisation are a strong indicator of early collapse, and therefore indicative of an {\it increased} central dark matter density. We parameterise this differentiated effect by considering $\textrm{M}_{\star,\mathrm{post}}/\textrm{M}_{\star,\mathrm{pre}}$, where the numerator and  denominator represent the stellar mass formed after and before $z\sim6.5$, respectively. 
Our study covers the halo mass range $10^9 < \textrm{M}_{200} < 10^{10} \ \mathrm{M}_\odot$ (stellar masses between $10^4 < \textrm{M}_{\star} < 10^8 \ \mathrm{M}_\odot$), spanning both ultra-faint and classical dwarfs. In this regime, $\textrm{M}_{\star,\mathrm{post}}/\textrm{M}_{\star,\mathrm{pre}}$ correlates almost perfectly with the central dark matter density at $z=0$, even when including simulations with a substantially different variant of feedback and cooling. We provide fitting formulae to describe the newfound dependence.

\end{abstract}




\section{Introduction} \label{sec:intro}
Dwarf galaxies can offer insights into the particle nature of dark matter. Their abundance is indicative of the small-scale cosmological power spectrum and can place constraints on scenarios such as warm dark matter \citep[e.g.][]{Schneider12, Newton_2021}.
Their central dark matter densities, which can be estimated from kinematic analyses \citep[e.g.][]{Binney_1982, Adams_2014, Zoutendijk_2021}, place limits on the strength of self-interacting dark matter (e.g. \citealt{Yoshida_2000, Rocha_2013}; see also \citealt{Tulin_2018} for a review) or on the mass of fuzzy dark matter candidates \citep[e.g.][]{Hu00, Schive_2014, Mina_2022}.
However, the effect of feedback from stellar populations is also particularly strong in these low-mass galaxies due to their shallow potential wells. This baryonic imprint is hard to calculate and substantially complicates the interpretation of observations \citep[e.g.][]{Dekel_1986, Efstathiou_1992,  Efstathiou_2000, Benson_2002, Pontzen_2014, Bullock_2017}. 

Simulations have established that baryonic physics plays an important role in shaping the density profiles of low-mass dark matter haloes \citep[e.g.][]{Read_2005, Mashchenko_2008, Governato_2010, Pontzen_2012, Onorbe_2015, Benitez_Llambay_2019}.
The `dark matter heating' model (see \citealt{Pontzen_2014} for a review) suggests that potential fluctuations caused by the supernovae explosions and other feedback-driven outflows can inject energy into the dark matter particles, causing them to migrate outwards. This process therefore turns the dense dark matter `cusps' expected from pure cold dark matter simulations \citep[e.g.][]{Navarro_1996, Read_2005, Teyssier_2013, DiCintio14, Read_2016} into `cores' of near-constant density. While a single blow-out cannot transform cusps into cores \citep[][]{Gnedin_2002}, repeated fluctuations from cycles of gas cooling and feedback can disrupt the dark matter particles orbits, causing the halo to expand and eventually flattening the inner regions of the density profiles. The inner density can also be lowered through a combination of stellar feedback and dynamical friction heating from subhalos, gas clouds and/or star clusters (\citealt{Orkney_2021}; see also \citealt{El_Zant_2001, Nipoti_2014}).

The stellar-to-halo mass ratio ($\textrm{M}_{\star} / \textrm{M}_{200}$) is a primary determinant of the extent of baryonic core creation processes at $z=0$ \citep{Penarrubia_2012, Di_Cintio_2013, DiCintio14,Tollet16, Bouche_2022, Deleo_2023}. The ratio $\textrm{M}_{\star} / \textrm{M}_{200}$ itself varies systematically with mass $\textrm{M}_{200}$ \citep[e.g.][]{Behroozi_2010, Moster_2013, Read_2017, Garrison-Kimmel_2017, Nadler_2020, Munshi_2021, Danieli_2023}. 
In smaller haloes, with low $\textrm{M}_{\star} / \textrm{M}_{200}$, stellar feedback is too weak to significantly alter the trajectories of the dark matter particles, resulting in cuspy inner density profiles. 
As the stellar mass increases, feedback becomes sufficiently strong to turn dark matter cusps into cores \citep{Penarrubia_2012, Read_2019}. 
As the $\textrm{M}_{\star} / \textrm{M}_{200}$ ratio increases further, the pressurisation of the interstellar medium becomes high enough that supernovae cannot generate coherent potential fluctuations across the galaxy as a whole, making dark matter heating inefficient once more.
This picture has been shown to provide qualitative agreement with observations of real galaxies for $-3 \lesssim \mathrm{log}(\textrm{M}_{\star} / \textrm{M}_{200}) \lesssim -1.75$, although the uncertainties in the data remain high \citep{Read_2019, Bouche_2022, Cooke_2022, Collins_2022}. 

However, there is no theoretical guarantee that $\textrm{M}_{\star} / \textrm{M}_{200}$ is the \textit{sole} determinant of core size and shape. From simulations, \citet{Chan_2015} noted that star formation that is contemporaneous with rapid halo collapse is inefficient at forming cores in simulations.  
Observationally, \citet{Oman_2015} highlight significant diversity in dwarf irregular galaxy rotation curves at fixed stellar mass. While this may in part reflect observational biases from non-circular motions or holes driven by stellar feedback in the HI gas distributions \citep[][]{Read_2016, Oman_2018, Marasco_2018, Roper_2023},  
diversity in the observed population underscores the need for theoretical analyses that explore factors other than total stellar mass.   
This need is the focus of our investigation.

In this work we analyse dark matter density profiles from the Engineering Dwarfs at Galaxy formation's Edge (\textsc{edge}) suite of state-of-the-art radiation-hydrodynamical simulations of dwarf galaxies \citep{Agertz2020EDGE}. We include simulations from both the \textsc{edge1} and \textsc{edge2} suite, the latter of which has updated physics that we will describe shortly. 
To study the impact of baryonic feedback, we focus primarily on the central halo density measured at $r=150\,\mathrm{pc}$, a radius at which observational density estimates may be made both in dwarf spheroidal and irregular galaxies \citep[][]{Read_2019}. The cusp-core transformation process acts to lower this density. 
We analyse the main factors that drive the core creation, showing that the {\it timing} of star formation has an important bearing, one which is not captured by $\textrm{M}_{\star} / \textrm{M}_{200}$. This is in agreement with the basic physics of dark matter heating through potential fluctuations, as discussed above, but has hitherto received limited attention.

This paper is organised as follows. Section \ref{sec:simulations} gives an overview of the \textsc{edge2} simulations physics and how it differs from the previous simulation suite (\textsc{edge1}), and an outline of the analysis pipeline. In Section \ref{sec:results} we present our key results and motivate the need for a new quantity which can represent the effect of feedback on \textsc{edge} haloes better than the widely used stellar-to-halo mass ratio ($\textrm{M}_{\star} / \textrm{M}_{200}$). 
In Section~\ref{sec:discussion} we present our conclusions, and discuss future theoretical and observational prospects.

\section{Simulations} \label{sec:simulations}
\textsc{edge}2 is a state-of-the-art suite of radiation-hydrodynamical dwarf galaxy simulations. 
The suite covers the mass range $10^9 < \mathrm{M_{200}} < 10^{10} \ \mathrm{M_{\odot}}$ (corresponding to $10^4 < \textrm{M}_{\star} < 10^8 \ \mathrm{M_{\odot}}$ in stellar masses).
The numerical resolution is the same as in the previous \textsc{edge}1 runs (dark matter particle mass $m_{\mathrm{DM}} = 940 \ \Msol$, and maximum spatial resolution in hydrodynamics and gravity of $\Delta x \approx 3$ pc), but the original galaxy formation model (described in detail in \citealt{Agertz2020EDGE, Rey2020}) has been significantly updated. 

Some of the most notable changes introduced in \textsc{edge}2 are:
\begin{enumerate}
  \item a new implementation of non-equilibrium cooling which follows \citet{Rosdahl2013}, modified from the equilibrium cooling from \citet{Courty2004} used in \textsc{edge}1;
  \item the inclusion of photo-ionization feedback using radiative transfer following the setup described in \citet{Agertz2020EDGE} across all the simulated haloes;
  \item an update of the UV background from a modified \citet{Haardt1996} model (see \citealt{Rey2020} for a discussion) to the \citet{Faucher-Giguere2020} photo-ionization and photo-heating rates.
\end{enumerate}
A more detailed description of all the changes introduced in the new \textsc{edge}2 simulations will be provided in an upcoming paper (Rey et al., in prep).

In this paper, we look at 8 isolated dark matter haloes taken from the \textsc{edge}2 simulations, each hosting a dwarf galaxy. Six of the eight haloes are re-runs of \textsc{edge}1 objects\footnote{These are Halo1445, 1459, 600, 605, 624, 383. The complete suite was first introduced in \citet{Orkney_2021, Rey_2022}; and \citet{gray_2024}.}, and can therefore be used as direct comparisons to study the effects of different feedback recipes. Halo261 and 339 are new to the \textsc{edge}2 simulation suite.

The dwarf galaxies are all evolved in isolated cosmological environments (i.e. they are not satellites). 
The high resolution allows us to model supernovae (SN) explosions as discrete events and to track their effects on the galaxies explicitly, removing the uncertainties of modelling these processes as sub-grid physics.
Our galaxies are evolved to $z=0$ from cosmological zoomed initial conditions constructed with the \textsc{genetIC} software (\citealt{Stopyra2021}).
All our simulations use the parameters $\Omega_{m} = 0.3086$, $\Omega_{\Lambda} = 0.6914$, $h = 0.6777$, $\sigma_8 = 0.8288$, $n_s = 0.9611$ which are compatible with \cite{PlanckCollaboration2014}. 
We follow the evolution of dark matter, stars, gas and radiation using the adaptative mesh refinement hydrodynamics code \textsc{ramses-rt} (\citealt{Teyssier2002, Rosdahl2013}). 
The star formation criteria in \textsc{edge}2 remain unchanged from \textsc{edge}1.

We process each \textsc{edge}2 simulation with the \textsc{adaptahop} halo and subhalo finder (\citealt{Aubert2004, Tweed2009}). 
\textsc{edge}1 simulations are processed with the \textsc{hop} halo finder (with parameters as described in \citealt{Eisenstein1998}). Because we study the deep interior of the major progenitor of a single galaxy in each simulation, we do not expect the difference in halo finders to have any bearing on our results. We match haloes and subhaloes between simulation snapshots to build merger trees using the \textsc{pynbody} (\citealt{Pontzen2013}) and \textsc{tangos} (\citealt{Pontzen2018}) libraries, in an identical way between the two \textsc{edge} versions.
Halo centres are identified using the shrinking-sphere algorithm (\citealt{Power2003}), and densities estimated by binning particles. We use 100 log-spaced bins of width ranging approximately from $2 \times 10^{-3}$ to $2.9$ kpc, and smooth them over a moving window of $20$ bins. Densities are shown exterior to 30 pc, which is sufficient to ensure that numerical softening effects are strongly subdominant \citep[][]{Muni_2023}. 

Where we show errors on densities, these are found by calculating the Poisson noise in each bin then using standard Gaussian error propagation to estimate an error on the moving-average quantities; however these statistical errors are typically small since our simulations are well resolved. While in the future we will use improved dynamics-based density estimators \citep[][]{Muni_2023}, for this initial work the statistical errors are strongly subdominant to the trends of interest.

\section{Results} \label{sec:results}

\subsection{Stellar-to-halo mass relation}

\begin{figure}
     \centering
      \includegraphics[width=\columnwidth]{{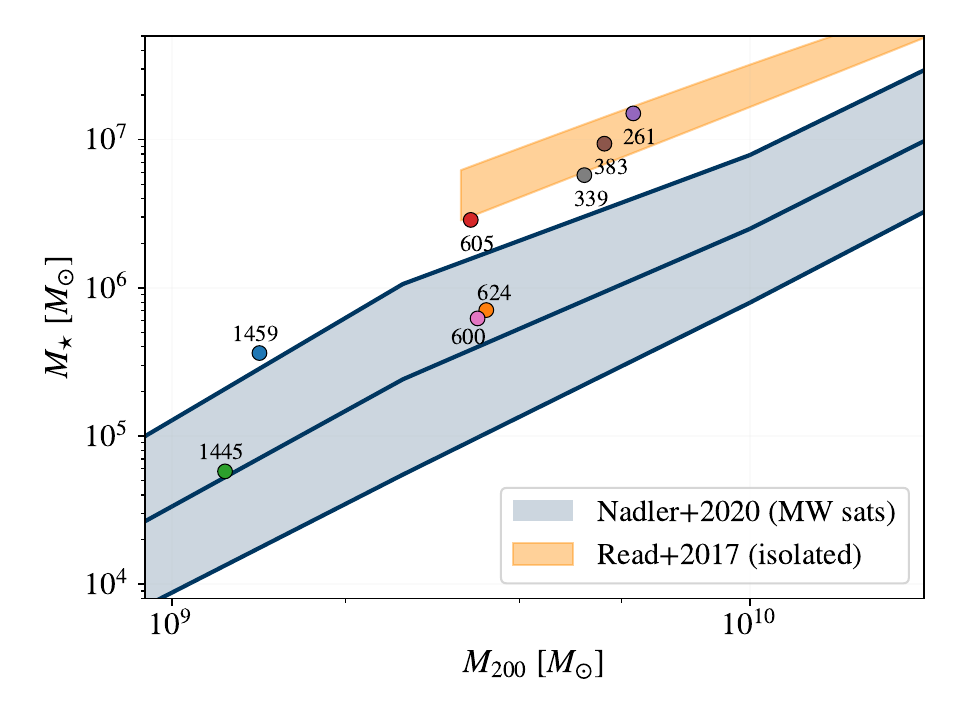}}
     \caption{Stellar-mass halo-mass (SMHM) relation of the \textsc{edge}2 haloes. 
     Abundance matching relations from real dwarfs are shown in blue and orange.
     The blue shading shows the 16-84\% confidence interval for Milky-Way satellites from \citet{Nadler_2020}. The orange shading is obtained from isolated dwarfs in \citet{Read_2017}. The \textsc{edge}2 results agree well with the Milky-Way satellites relation for small values of $\textrm{M}_{200}$ where the star formation rate is predominantly regulated by reionisation quenching. As the halo mass grows, environmental effect become significant and the \textsc{edge}2 results show better agreement with the relation from isolated dwarfs.}
     \label{figure_Mstar_Mhalo_new}
\end{figure}

Fig.~\ref{figure_Mstar_Mhalo_new} shows the stellar-mass halo-mass (SMHM) relation measured for the \textsc{edge}2 haloes. We take $\textrm{M}_{200}$ to be the mass enclosed by $\textrm{r}_{\mathrm{200}}$, the radius within which the mean density is 200 times higher than the critical density. 
We can approximately divide our \textsc{edge}2 haloes into three categories: 1445 and 1459 with lower halo masses ($\sim 10^{9}\,\Msol$); 600, 605, and 624 with intermediate halo masses ($\sim 3 \times 10^{9} \Msol$); and 339, 383, 261 with larger $\textrm{M}_{200} (\sim 4 \times 10^9 \Msol$). We notice that even when the haloes have similar $\textrm{M}_{200}$, their $\textrm{M}_{\star}$ can vary significantly, showing a dependence of stellar mass on the formation history \citep[see][]{Rey19}.

The bands in Fig.~\ref{figure_Mstar_Mhalo_new} allow us to compare our results with two empirical relations obtained from abundance matching in observed dwarf galaxies: \citet{Read_2017}, which refers to isolated dwarf galaxies similar to our simulated objects, but only covers the high-mass range of our sample; and \citet{Nadler_2020}, which spans the correct mass range for our haloes, but is obtained from Milky-Way satellites data.

In the low- and intermediate-mass regime, the agreement between the \textsc{edge}2 data and the Milky Way satellites relation is good; our stellar masses are at the upper end of the observationally inferred range, but this can likely be explained by our galaxies being isolated rather than satellites \citep[although see][]{Christensen_2024}. The duration of star formation of ultra-faint dwarfs over time is mostly controlled by reionisation quenching, but 
in MW satellites one would expect a secondary environmental effect from tides, stripping and other disruption \citep[e.g.][]{Shipp_2018, Mutlu_Pakdil_2019, Weerasooriya_2023}. 
In the larger-mass regime, these environmental effects may be a dominant factor in the MW satellite population since reionisation does not quench them \citep[e.g.][]{Fattahi_2018, Pace_2022}. 
In these cases, our simulations form too many stars to be compatible with  MW satellite observations, but are in excellent agreement with the relation from isolated galaxies. See Rey et al. (in prep) for a more detailed discussion on the SMHM relation in \textsc{edge}2, and \citet{Kim_2024} for a semi-analytic code calibrated to predict \textsc{edge} stellar masses across a much larger statistical sample of dwarf galaxies.

Our simulations are not specifically tuned to reproduce the SMHM relation, and as such the comparison above gives some confidence. Nonetheless, it should be viewed with caution.  
\citet{Nadler_2020} advise that their model of the Milky Way satellite SMHM is a useful empirical relation, but is not intended to capture the complexities of star formation in ultra-faint dwarfs, where there is a very strong sensitivity to accretion history \citep{Rey19, Kim_2024}. 
The behaviour of the \citet{Read_2017} field relation for $\textrm{M}_{\star} \lesssim 4 \times 10^6 \ \Msol$ is also unclear since a direct extrapolation to lower halo masses is not necessarily justified. Also there is evidence that this relation is based on HI-rich dwarfs which are outliers in stellar mass for their halo mass \citep[][]{Kim_2024}. 
In this context, we conclude that the \textsc{edge}2 results approximately span the correct stellar mass range given their halo masses, but there are still considerable uncertainties in how galaxies truly occupy the SMHM plane; future facilities such as the Vera Rubin Observatory will have a major impact on tightening these uncertainties \citep[][]{Simon_2019}.

\subsection{Core creation as a function of $\mathbf{\Mstar / M_{200}}$} \label{subsec:failure_Mstar_M200}

\begin{figure}
     \centering
      \includegraphics[width=\columnwidth]{{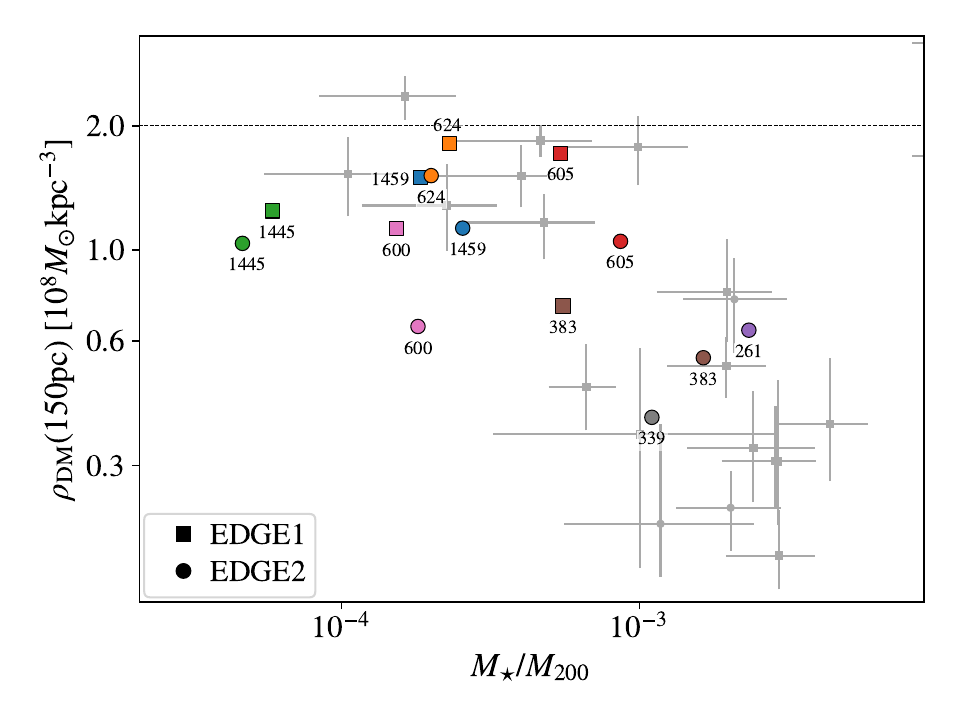}}
     \caption{Dark matter density at $r=150$pc as a function of the stellar-to-halo mass ratio in the \textsc{edge}2 and \textsc{edge}1 haloes, shown as circles and squares respectively. The \textsc{edge}2 simulations have an updated physics model compared to \textsc{edge}1, as discussed in Section \ref{sec:simulations}. The Poisson errors on the dark matter densities are small (of the order a few per cent) and therefore not visible in the figure. The light grey points represent dwarf galaxies data from \citet{Read_2019}. The horizontal dashed line shows the expected inner density for a $\Lambda$CDM cosmology at the mass scale of the \textsc{edge} haloes. The \textsc{edge} results show qualitative agreement with the data, however there is considerable scatter both in the simulations and the observations (although the latter has large statistical uncertainties). This suggests that the $\textrm{M}_{\star}/\textrm{M}_{200}$ ratio alone is not sufficient to predict central dark matter density profiles.} 
     \label{figure_data_Mstar_M200_comparison}
\end{figure}

Fig.~\ref{figure_data_Mstar_M200_comparison} shows the dark matter density at $r=150$ pc as a function of the stellar-to-halo mass ratio for our \textsc{edge} haloes. The observational significance of this radius was discussed in Section \ref{sec:intro}; from now on we will refer for brevity to $\rho_{\mathrm{DM}}$ implying that it has been evaluated at this radius, unless otherwise specified.  A higher value of the density $\rho_{\mathrm{DM}}$ corresponds to `cuspier' haloes. The grey points show the values measured for real dwarf galaxies (both isolated and satellites) obtained from \citet{Read_2019} (and references therein). Our simulation suite covers halo masses up to $\sim 10^{10} \Msol$, allowing us to probe only the low-mass end of this relation. The simulation densities and associated statistical uncertainties were calculated as described in Section~\ref{sec:simulations}, and the resulting Poisson errors are too small to display (ranging between $1.8\times 10^6$ and $3.5\times 10^6\,\Msol\,\mathrm{kpc}^{-3}$, i.e. of the order of a few percent).

\begin{figure*}
     \centering
      \includegraphics[width=\textwidth]{{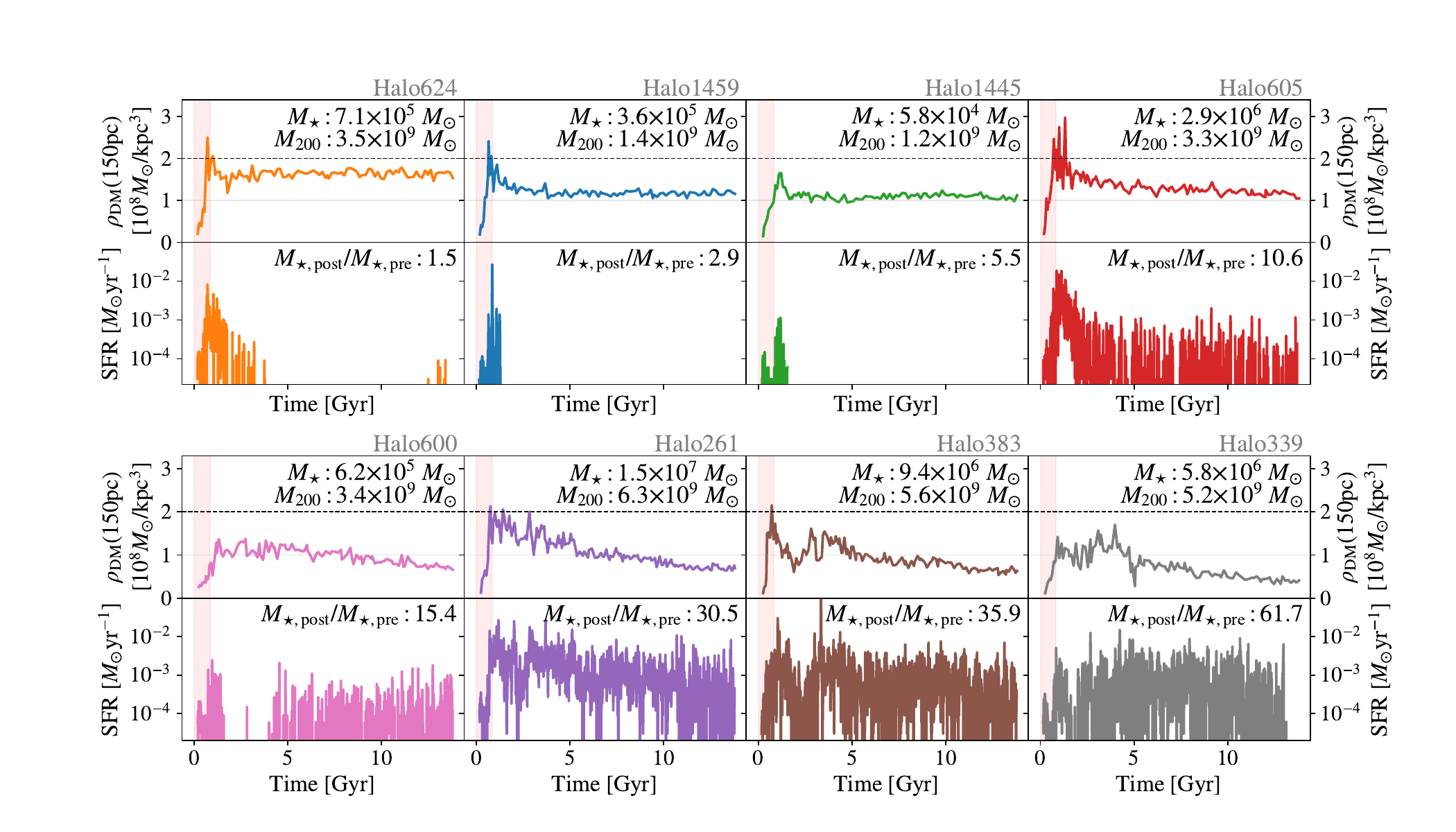}}
     \caption{Evolution of the dark matter densities at 150 pc for the 8 \textsc{edge}2 haloes (top rows) alongside their galaxies' star formation histories (bottom rows) as a function of cosmic time since the Big Bang. The pink shaded region shows the time prior to reionisation ($z \lesssim 6.5$) and the horizontal dashed line at $\rho_{\rm DM} = 2 \times 10^8\,M_{\odot}/{\rm kpc}^3$ shows the expected density for a $\Lambda$CDM cosmology (same as in Fig. \ref{figure_data_Mstar_M200_comparison}). The haloes are ordered according to their $\textrm{M}_{\star, \mathrm{post}}/ \textrm{M}_{{\star, \mathrm{pre}}}$ ratio (the stellar mass post-reionisation divided by the stellar mass prior to reionisation). The first three haloes from the top retain cuspy profiles, while the rest of the haloes form cores of various sizes. Haloes which show star formation continuing until today display lower central dark matter densities; conversely, haloes which have early star formation reach higher concentrations before reionisation, and therefore have higher inner densities at the present day.} 
     \label{figure_density_over_time}
\end{figure*}

The results of the \textsc{edge}2 simulations (circles) show broad agreement with the data, with an average decrease in the central densities as the value of $\textrm{M}_{\star} / \textrm{M}_{200}$ increases, consistent also with the theoretical picture discussed in Section~\ref{sec:intro}. 
However, there is significant scatter in the simulation results. The Pearson's coefficient for the combined \textsc{EDGE}1 and \textsc{EDGE}2 suites of simulations was found to be $r = -0.54$. 
Comparing pairs of haloes in \textsc{edge}2 such as Halo339 and 605, or Halo600 and 624, we obtain similar values of the $\textrm{M}_{\star} / \textrm{M}_{200}$ ratio, but diverse central dark matter densities, with variations up to a factor of 2.7.
Conversely, Halo600, 383, and 261 all have similar values of $\rho_{\mathrm{DM}}$, but their stellar-to-halo mass ratio is extremely varied.
The observational data is also compatible with these variations in  $\rho_{\mathrm{DM}}$ at fixed $\textrm{M}_{\star}/\textrm{M}_{200}$, although the small sample size and large statistical uncertainties make it difficult to assess the population scatter quantitatively.

While the correlation of $\rho_{\mathrm{DM}}$ with $\textrm{M}_{\star}/\textrm{M}_{200}$ is clear in the \textsc{edge}2 simulations ($r = -0.58$) and the combined \textsc{edge}1+2 sample, if one removes the \textsc{edge}2 simulations (circles) and considers  \textsc{edge}1 (squares) alone, there is no identifiable trend ($r = -0.13$). For example, Halo383 and 600 in \textsc{edge}1 both have high values of $\textrm{M}_{\star}/\textrm{M}_{200}$ but the former is cored and the latter cuspy.
Furthermore Halo600 and 605 are cored in \textsc{edge}2 but retain high central densities in \textsc{edge}1, underscoring the impact of changing the feedback implementation (Section \ref{sec:simulations}). 

From the above it appears that the $\textrm{M}_{\star}/\textrm{M}_{200}$ ratio alone is not a reliable indicator of the presence of cores in the \textsc{edge} haloes. 
The ratio does not capture essential information about the history of the galaxies which affects the dark matter density at present day. We will next show that these results, including the differences between feedback recipes, can be explained once the timing of star formation is taken into account.

\subsection{The importance of star formation timing}

Fig.~\ref{figure_density_over_time} shows the evolution over time of the dark matter density $\rho_{\mathrm{DM}}$ at our chosen radius ($r=150$ pc) for the 8 \textsc{edge}2 haloes (top rows) alongside the galaxies' star formation histories (bottom rows), as a function of cosmic time since the Big Bang.  The small fluctuations in $\rho_{\mathrm{DM}}$ arise from statistical noise due to the limited number of particles. 

Halo1459, 624, and 1445 (first three in the top row) are the only haloes whose $\rho_{\mathrm{DM}}$ remain constant as time progresses, meaning they retain steep cusps throughout their history. The inner gradients of the remaining haloes flatten over time forming cores of various sizes\footnote{Note that the value of the density at 150 pc alone does not strictly indicate the presence of a cusp or core at $z=0$. For a clearer understanding of why we refer to Halo1445, 1459, 624 as cuspy and to the rest of the haloes in \textsc{edge}2 as cored, see the density profiles in Fig. \ref{core_NFW_fit}.}. 
The decrease in the central density of the cored haloes is gradual, underscoring that stellar feedback causes the dark matter halo to slowly expand over time through several smaller outflow episodes, rather than a single explosive incident.\footnote{We do not observe any cores forming and then reverting back into cusps at later times; in the haloes where the gradient does not gradually flatten over time, the density remains constant over their entire history. This may be due to the limited sample size (see Section~\ref{sec:discussion}). }

Among the cored haloes, we notice a correlation between their central densities and their virial masses (which is given in each panel), with more massive haloes showing progressively lower $\rho_{\mathrm{DM}}$ on average. 
The decrease in $\rho_{\mathrm{DM}}$ with $M_{200}$ is consistent with the average decline in densities as a function of $\textrm{M}_{\star}/\textrm{M}_{200}$ in Fig.~\ref{figure_data_Mstar_M200_comparison}, since $M_{200}$ and $M_*$ are connected by the SMHM relation.
However, as noted above for the stellar-to-halo mass ratio, there is considerable scatter around this trend. Taking, for example, the comparison between Halo600 and Halo624, both have  almost identical $\textrm{M}_{200}$ and $\textrm{M}_{\star}$ values but their final dark matter density is very different.
The reason for this becomes clear when we compare the star formation historiess (SFHs; bottom rows of Fig.~\ref{figure_density_over_time}). While Halo624 forms almost all its stars at high redshift, many of them before reionisation, Halo600 builds up its stellar population predominantly at late times. As such, it proceeds through many more cycles of gas collapse, star formation and gas expulsion; this cyclical engine is necessary to gradually reduce the dark matter density over time \citep{Pontzen_2014}. Moreover, because Halo624 collapses quickly at high redshift, its central cusp is more concentrated at early times, reaching a peak value of $\rho_{\mathrm{DM}} = 2.5 \times 10^8 \ \Msol \mathrm{kpc}^{-3}$ compared with Halo600's $\rho_{\mathrm{DM}} = 1.4 \times 10^8 \ \Msol \mathrm{kpc}^{-3}$. 

The link between earlier star formation and higher dark matter densities is further reinforced by comparing Halo600 with Halo605. Both objects have similar virial masses and extended star formation after reionisation, yet the final density of Halo605 is $\rho_{\mathrm{DM}} = 1.0 \times 10^8 \ \Msol \mathrm{kpc}^{-3}$, compared with the lower $\rho_{\mathrm{DM}} = 6.6 \times 10^7 \ \Msol \mathrm{kpc}^{-3}$ of Halo600. This is despite Halo605 forming {\it more} stars ($2.9 \times 10^6\,\Msol$) than Halo600 ($0.6 \times 10^6\,\Msol$); the central density trend is inverted with respect to expectations from $\textrm{M}_{\star}/\textrm{M}_{200}$. The inverted trend arises because Halo605 collapsed earlier, forming with a higher dark matter density than Halo600 at early times ($t\sim 2\,\mathrm{Gyr}$); consequently, the $z=0$ dark matter density of the former remains high compared with the latter. This difference is correlated with early star formation, but not causally affected by it; we find that it is also present in the dark-matter-only simulations, where the densities at $t\simeq 2\,\mathrm{Gyr}$ are $1.4 \times 10^8 \Msol \mathrm{kpc}^{-3}$ for Halo600 and $2.6 \times 10^8 \Msol \mathrm{kpc}^{-3}$ for Halo605.

In summary, there are two separate mechanisms that link dark matter density and star formation history. The difference between Halo600 and Halo624 can be causally explained by the amount of sustained, {\it late}-time star formation. A greater amount of sustained star formation leads directly to lower central densities, through the gravitational dynamics discussed in Section \ref{sec:intro}. Meanwhile, the difference between Halo600 and Halo605 can be explained by the high central concentration due to early collapse of the latter. {\it Early} star formation is correlated with higher central densities, since it is an indicator of early collapse and the feedback does not have time to significantly flatten the density profile.

This perspective also explains the densities in the remaining \textsc{edge}2 haloes. The cuspy haloes (first three in the top row of Fig.~\ref{figure_density_over_time}) have already formed almost all of their stars by $z \sim 6$, while cored haloes see extended star formation until today, reducing their central dark matter density steadily over the Hubble time. 
Among the cored haloes, we see a correlation between the dark matter density and amount of stars formed at recent times: lower densities at $z=0$ are linked to particularly active star formation post-reionisation. 

\label{sec:importance_Mpre}

While there are a variety of ways in which one could divide star formation into `early' and `late' regimes, reionisation provides a natural transition time. As mentioned in Section \ref{sec:intro}, reionisation completely quenches low mass dwarf galaxies. The smallest ultra-faint dwarf galaxies form all their stars at high redshift, before losing their gas. The most massive dwarfs, by contrast, have a sufficient potential depth and reservoir of dense gas by the time of reionisation that they self-shield against the external ionising radiation. In such cases, internal self-regulation is the primary factor limiting star formation rate, and reionisation has only a secondary impact. In intermediate-mass cases, the timing of star formation is strongly dependent on the accretion history relative to reionisation, and in some cases the galaxy may temporarily quench before later achieving a sufficient mass to restart star formation \citep{Rey19, Rey2020}. 

The pivotal role of reionisation in these cases underscores why it forms a natural division between the two competing effects on the dark matter density. For the rest of our investigation we will use 
\begin{equation}
    \mathrm{M}_{\star,\mathrm{post}} = \int^{t_\mathrm{today}}_{t=t_{\mathrm{reion}}}{\mathrm{SFR}} \ \mathrm{d}t,
\end{equation}
(i.e. the amount of star formation which occurred in the galaxy's progenitors after reionisation) as a proxy for the late-time star formation capable of flattening dark matter cusps. Conversely, we will use the quantity 
\begin{equation}
    \mathrm{M}_{\star, \mathrm{pre}} = \int^{t_\mathrm{reion}}_{t=0}{\mathrm{SFR}} \ \mathrm{d}t,
\end{equation}
(i.e. the amount of star formation which occurred within the galaxy's progenitors prior to reionisation) as a proxy for the early collapse which encourages higher densities of dark matter. While reionisation is a gradual process, we adopt $z\simeq6.5$ ($t \simeq 0.84$ Gyr) when dividing the stellar populations. In Fig.~\ref{figure_density_over_time}, as well as in the calculation of $\mathrm{M}_{\star, \mathrm{post}}$ and $\mathrm{M}_{\star, \mathrm{pre}}$, we use the total mass formed at birth in the relevant stellar populations, rather than their remaining mass at $z=0$, and we include stars formed in all progenitors of the galaxy (rather than just the major progenitor). 

In Section~\ref{sec:discussion} we will discuss these choices further, but our results below show that they successfully summarise a complex story in two simple competing effects. Specifically, we now explore the use of the $\mathrm{M}_{\star,\mathrm{post}} / \mathrm{M}_{\star,\mathrm{pre}}$ ratio as an indicator for the expected central density of CDM haloes.

\subsection{The $\mathbf{M_{\star,\mathrm{\mathbf{post}}} / M_{\star,\mathrm{\mathbf{pre}}}}$ ratio} 
\label{sec:post-pre-ratio}

\begin{figure}
     \centering
      \includegraphics[width=\columnwidth]{{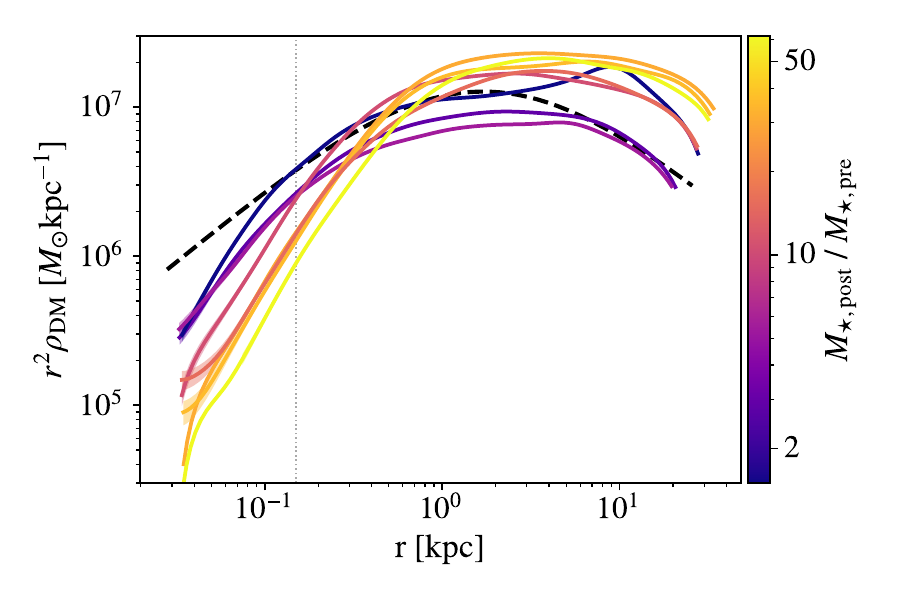}}
     \caption{Dark matter density profiles of the \textsc{edge}2 haloes at $z=0$, colour-coded by the value of their $\mathrm{M}_{\star,\mathrm{post}}/\mathrm{M}_{\star,\mathrm{pre}}$ ratio (the post-reionisation stellar mass as a fraction of the stellar mass prior to reionisation). The $y$-axis has been multiplied by $r^2$ to emphasize the inner regions. The NFW profile (dashed black line) for a dark matter halo of mass $2 \times 10^8 \Msol$ and concentration $c=15$ at $z=0$ is shown for comparison. The vertical dotted line marks $r = 150$pc. The shapes of the dark matter density profiles clearly depend on the galaxies' values of the $\mathrm{M}_{\star,\mathrm{post}}/\mathrm{M}_{\star,\mathrm{pre}}$ ratio, especially in the inner regions.} 
     \label{figure_profiles}
\end{figure}

In Fig.~\ref{figure_profiles} we show the dark matter density profiles of the \textsc{edge}2 haloes. The $y$-axis has been multiplied by $r^2$ to emphasize the shape of the inner regions, and the profiles have been colour-coded based on the values of the galaxies' $\mathrm{M}_{\star,\mathrm{post}}/\mathrm{M}_{\star,\mathrm{pre}}$ ratio. Lighter colours indicate higher $\mathrm{M}_{\star,\mathrm{post}}/\mathrm{M}_{\star,\mathrm{pre}}$, i.e. star formation occurring at more recent times. The $\mathrm{M}_{\star,\mathrm{post}}/\mathrm{M}_{\star,\mathrm{pre}}$ ratio accurately correlates with the extent of core formation in the haloes: the higher the value, the lower the central density relative to the unmodified NFW profile (black dashed line).

\begin{figure}
     \centering
      \includegraphics[width=\columnwidth]{{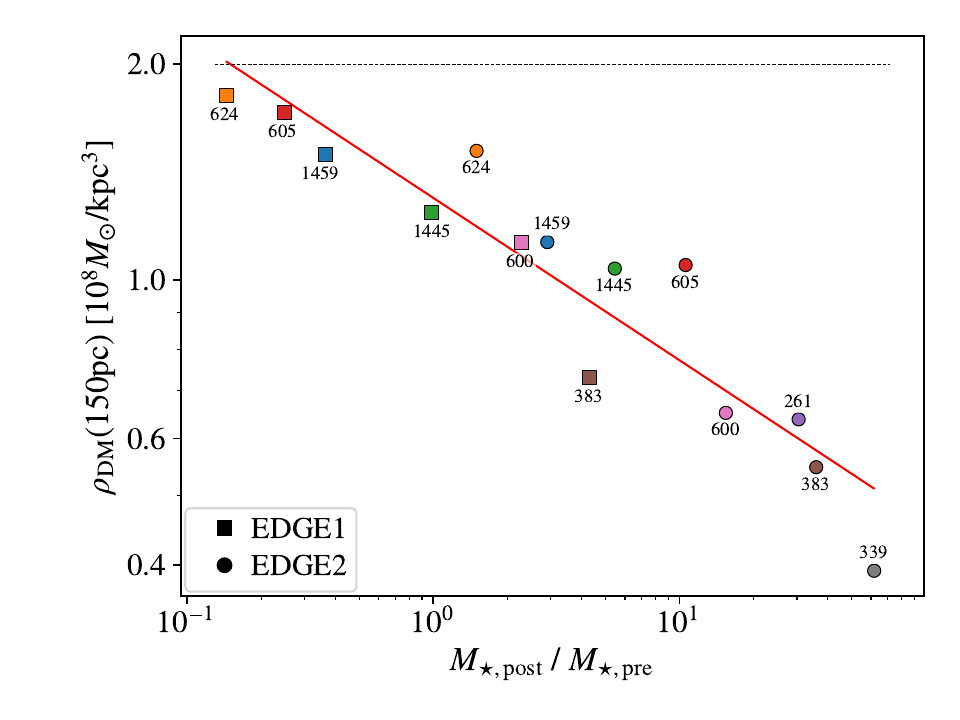}}\vspace{-0.5cm}
     \caption{Dark matter density at 150pc as a function of the $\mathrm{M}_{\star, \mathrm{post}} / \mathrm{M}_{\star, \mathrm{pre}}$ ratio. The circles show the \textsc{edge}2 haloes, while the squares are results from the \textsc{edge}1 simulations. A fit to the combined relation, given by Eq.~\eqref{eq:rhodm-fit}, is shown in red. The horizontal dashed line shows the expected inner density from $\Lambda$CDM cosmology at the mass scale of the \textsc{edge} haloes. We notice a tight correspondence between the central dark matter density at $z=0$ and $\mathrm{M}_{\star, \mathrm{post}} / \mathrm{M}_{\star, \mathrm{pre}}$ which was not captured by the $\textrm{M}_{\star}/\textrm{M}_{200}$ ratio.} 
     \label{figure_Mpost_M200_fit}
\end{figure}

\begin{figure*}
     \centering
      \includegraphics[width=\textwidth]{{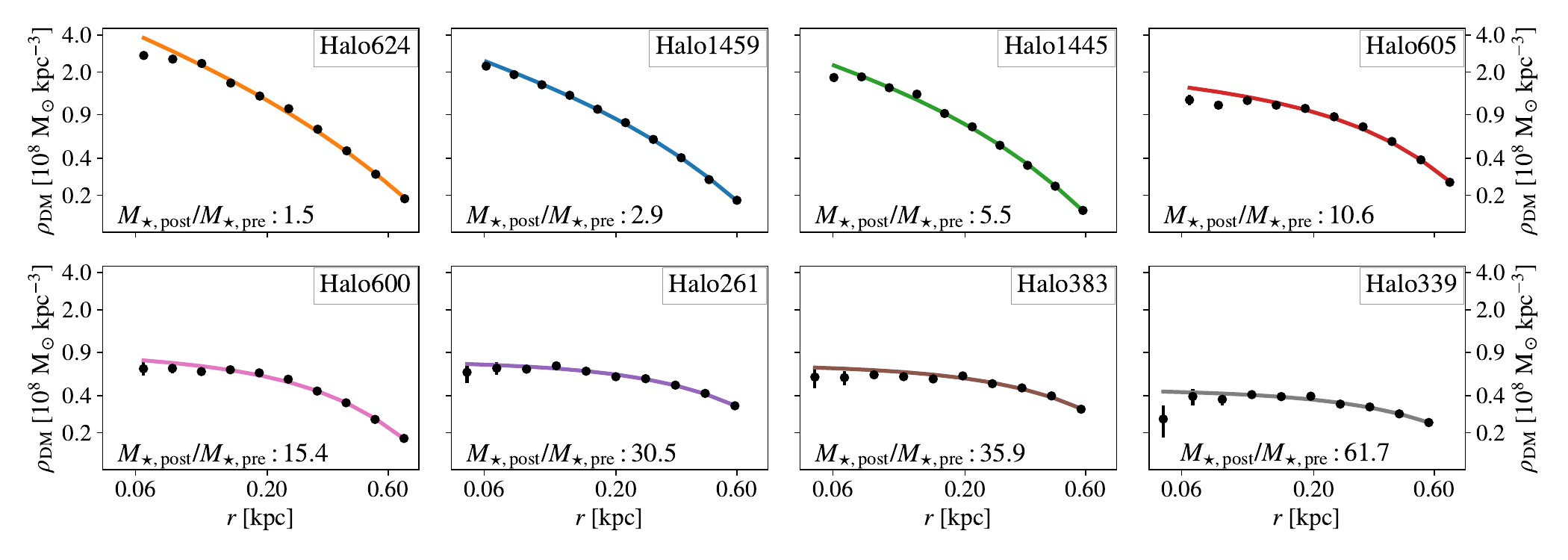}}\vspace{-0.7cm}
     \caption{Fit to the density profiles of the \textsc{edge}2 haloes using the coreNFW profile. The sizes of the cores are set to be directly proportional to the haloes' half-light radii, and we fit the parameter $n$ which ranges from an unmodified cusp ($n=0$) or flat core ($n=1$). The profiles were cut between $r=0.05$ kpc and $0.7$ kpc to capture the variation in the dark matter central regions across haloes.}
     \label{core_NFW_fit}
\end{figure*}

Fig.~\ref{figure_Mpost_M200_fit} shows more quantitatively how  $\mathrm{M}_{\star,\mathrm{post}}/\mathrm{M}_{\star,\mathrm{pre}}$ is predictive of the central density evaluated at a fixed $150\,\mathrm{pc}$ radius.
The correlation of this dark matter inner density is much stronger with $\mathrm{M}_{\star,\mathrm{post}}/\mathrm{M}_{\star,\mathrm{pre}}$ than with the stellar-to-halo mass ratio (Fig.~\ref{figure_data_Mstar_M200_comparison}), especially for the \textsc{edge}1 haloes (squares). If we fit these points with the relation
\begin{equation}
\log_{10} \rho_{\mathrm{DM}}(150\,\mathrm{pc}) = m \log_{10}\left(\frac{\mathrm{M}_{\star,\mathrm{post}}}{\mathrm{M}_{\star,\mathrm{pre}}}\right) + c\,,\label{eq:rhodm-fit}
\end{equation}
the values of the coefficients for the combined sample are $m=-0.23 \pm 0.03$ and $c = 8.13 \pm 0.03$. This fit is shown as a red line in Fig.~\ref{figure_Mpost_M200_fit}.

As examples of why the correlation has improved, consider the case of \textsc{edge}2 Halo600 contrasted with Halo383 and 261; as discussed in Section \ref{subsec:failure_Mstar_M200}, Halo600 has a very different $\textrm{M}_{\star} / \textrm{M}_{200}$, but a similar value of $\rho_{\mathrm{DM}}(150$pc). Fig.~\ref{figure_Mpost_M200_fit} highlights that Halo600 has an $\mathrm{M}_{\star,\mathrm{post}}/\mathrm{M}_{\star,\mathrm{pre}}$ ratio in approximate agreement with Halo383 and 261, making sense of the low central density. 
Similarly, Halo339 and 605 have comparable $\textrm{M}_{\star}/\textrm{M}_{200}$, but Halo605 underwent most of its star formation pre-reionisation, resulting in a lower value of $\mathrm{M}_{\star,\mathrm{post}}/\mathrm{M}_{\star,\mathrm{pre}}$ and higher central dark matter density. In other words, what appeared as random scatter in Fig.~\ref{figure_data_Mstar_M200_comparison} now appears as part of a tight trend in Fig.~\ref{figure_Mpost_M200_fit}.
We conclude that the scatter in the relationship between $\textrm{M}_{\star}/\textrm{M}_{200}$ and $\rho_{\mathrm{DM}}$ can be accounted for by the timing effects identified in Section~\ref{sec:importance_Mpre}. 


Since the duration of star formation in dwarf galaxies is heavily controlled by reionisation, our results are sensitive to the choice of separation time between pre- and post-reionisation stars, especially for the haloes where $\mathrm{M}_{\star,\mathrm{post}}$ is small. The dividing time we adopt ($t=0.84$ Gyr) maximises the Pearson's coefficient ($r=-0.93$) of the $\rho_{\mathrm{DM}} - \mathrm{M}_{\star,\mathrm{post}}/\mathrm{M}_{\star,\mathrm{pre}}$ relation for the combined suites of \textsc{edge}1 and \textsc{edge}2 simulations. This coefficient has a significantly increased magnitude compared with the correlation coefficient between $\rho_{\mathrm{DM}}$ and  $\textrm{M}_{*}/\textrm{M}_{200}$ ($r=-0.58$, see Section \ref{subsec:failure_Mstar_M200}), indicating that the $\mathrm{M}_{\star,\mathrm{post}}/\mathrm{M}_{\star,\mathrm{pre}}$ ratio possesses a much stronger correlation to the halo central density. When moving the dividing time by $\pm$ 0.1 Gyr, the joint Pearson’s coefficient decreases by $7\%$, showing a clear preference for the adopted reionisation time across the combined suite. 
However, differences in the adopted UV background mean that reionisation happens at different times between \textsc{edge}1 and \textsc{edge}2.
In \textsc{edge}2 taken alone, we see the tightest correlation earlier in time ($t = 0.72$ Gyr; $r=-0.96$).
Thus, residual scatter in the \textsc{edge}2 results (circles in Fig.~\ref{figure_Mpost_M200_fit}) is largely accounted for by the fact we adopt a fixed time for reionisation across the combined suite.

\subsection{Dependence on the sub-grid model}

The extent of baryonic core creation is known to be strongly sensitive to the way in which star formation and feedback is implemented in simulations \citep{Pontzen_2012, Onorbe_2015, Chan_2015, Benitez_Llambay_2019}. 
The most important factor for enabling a significant response from the dark matter is that gas reaches densities comparable to the central dark matter cusp, and that a significant fraction of the gas mass repeatedly flows in and out of the central $\sim$ kpc of the galaxy \citep[e.g.][]{Read_2005, Pontzen_2012, Pontzen_2014}. This naturally occurs if stars form only from high density gas and there is sufficient numerical resolution to capture the multiphase interstellar medium \citep[][]{Pontzen_2012, Dutton_2019}. \citet{Teyssier_2013} highlighted that this results in `bursty' star formation, with a peak-to-trough ratio of 5 to 10 and a duty cycle comparable to the local dynamical time. This prediction has been borne out by the data for dwarf galaxy populations (e.g. \citealt{Kauffmann_2014, McQuinn_2015, Emami_2019} and for a review see \citealt{Collins_2022}).

As mentioned in Section \ref{sec:simulations}, some important changes were introduced in the physics of the new \textsc{edge}2 simulations compared to \textsc{edge}1.
Other than the timing of reionisation that we discussed in Section~\ref{sec:post-pre-ratio}, the most notable changes relate to the new implementation of non-equilibrium cooling and the inclusion of radiative transfer, which was absent in the majority of \textsc{edge}1 runs. These factors result in considerable differences in the galaxy properties, including the SFHs (see Appendix \ref{appendixA}) and will be discussed more fully in Rey et al. (in prep).

As a result, \textsc{edge}1 haloes end up with higher central densities than \textsc{edge}2 haloes. Prior to reionisation, the \textsc{edge}1 simulations have intense large-scale outflows driven by supernovae, which lead to a hot ($T > 10^6$ K) unbound circumgalactic medium. The radiative feedback adopted in \textsc{edge}2 reduces the local intensity of star formation, weakening these outflows and keeping the circumgalactic gas at lower temperatures ($T \sim 10^4$ K) \citep{Agertz2020EDGE}, enabling more sustained star formation. 
As shown in Fig.~\ref{figure_data_Mstar_M200_comparison}, the different feedback recipes and the changes in the non-equilibrium cooling strength lead to \textsc{edge}2 simulations occupying a significantly different locus in the $\textrm{M}_{\star}/\textrm{M}_{200}$ -- $\rho_{\mathrm{DM}}$ plane compared to \textsc{edge}1 simulations.

This difference between suites is almost entirely erased when the central density is seen as a function of $\mathrm{M}_{\star,\mathrm{post}}/\mathrm{M}_{\star,\mathrm{pre}}$ (Fig.~\ref{figure_Mpost_M200_fit}), since the new ratio more directly captures the factors determining the central dark matter density. 
In this picture, \textsc{edge}1 and \textsc{edge}2 core creation physics appears very similar, but statistically significant differences do remain. When fitting the suites separately to Eq.~\eqref{eq:rhodm-fit}, 
the values of the coefficients for \textsc{edge}2 with reionisation time $t=0.72$~Gyr are $m_{\mathrm{EDGE2}}=-0.30 \pm 0.04$ and $c_{\mathrm{EDGE2}}=8.32 \pm 0.06$, and for \textsc{edge}1 with reionisation time $t=0.84$~Gyr are $m_{\mathrm{EDGE1}}=-0.24 \pm 0.04$ and $c_{\mathrm{EDGE1}}=8.08 \pm 0.02$.

In terms of the overall core creation process, these differences are strongly subdominant to the more basic effect that $\mathrm{M}_{\star,\mathrm{post}} / \mathrm{M}_{\star,\mathrm{pre}}$ changes when moving from $\textsc{edge}$1 to $\textsc{edge}$2 physics. On average the \textsc{edge}1 haloes tend to have lower values of $\mathrm{M}_{\star,\mathrm{post}} / \mathrm{M}_{\star,\mathrm{pre}}$ than \textsc{edge}2, explaining why they also are less prone to forming cores. 

As such, the qualitative significance of $\mathrm{M}_{\star,\mathrm{post}} / \mathrm{M}_{\star,\mathrm{pre}}$ to core formation is preserved regardless of at least some details of the feedback recipe. However, given that \textsc{edge}1 and \textsc{edge}2 share much of their subgrid physics in common, it will be of great interest to test whether other simulations with a resolved interstellar medium \citep[e.g.][]{Jeon_2017, Wheeler_2019, Munshi_2021, Gutcke_2022} 
also predict a major role for $\mathrm{M}_{\star,\mathrm{post}} / \mathrm{M}_{\star,\mathrm{pre}}$ in core creation.

\subsection{Density profiles fit with coreNFW model}
Density profiles in which the central density has been reduced by baryonic feedback can be fit using a modified version of the classic Navarro-Frenk-White (NFW) profile \citep{Navarro_1997}. \citet{Read_2016} introduced the coreNFW density profile which, in addition to the NFW parameters (scale radius $r_s$ and density $\rho_0$) uses two additional variables to describe the core size ($r_c$) and inner density slope ($n$). The latter ranges from $1$ (for a completely flat core) to $0$ (for a cuspy NFW profile). See \citet{Read_2016}, especially their Eq.~(16), for a precise definition. 
Assuming that the core size is directly proportional to the stellar half-mass radius, \citet{Deleo_2023} provided a fit to the remaining parameter $n$ as a function of $\textrm{M}_{\star}/\textrm{M}_{200}$ based on observational constraints.

To summarise the core creation process seen in our \textsc{edge}2 simulation results, we adopt a similar approach.
We perform a fit for $n$ (the central density slope) of each dark matter density profile. We also fit the density profiles for $\textrm{M}_{200}$ and $c$, which control the parameters of the classic NFW profile.
We expect the value of $n$ to evolve from $n=0$ (an NFW cusp) in haloes with small values of $\mathrm{M}_{\star,\mathrm{post}} / \mathrm{M}_{\star,\mathrm{pre}}$ to $n=1$ (a flat core) when the feedback-driven outflows become important. 
There is a degree of degeneracy between $n$ and the core size parameter $r_c$; we break this by setting $r_c$ to be directly proportional to the haloes V-band 3D half-light radii. The constant of proportionality ($\eta=1.8$) was calculated by letting both $n$ and $\eta$ vary when fitting the central regions ($0.05 \ \mathrm{kpc} <r<1$ kpc) of the profile of Halo339, which has a prominent core, and was then held fixed for all other haloes. 
Fig.~\ref{core_NFW_fit} shows the individual fits of the inner regions of the density profiles for the \textsc{edge}2 haloes. 
The density profiles were cut between $r=0.05$ kpc and $0.7$ kpc in order to capture the variation of the profiles' inner regions.
The fits are qualitatively good and capture the correct central slopes and core sizes across all the haloes.

\begin{figure}
     \centering
      \includegraphics[width=\columnwidth]{{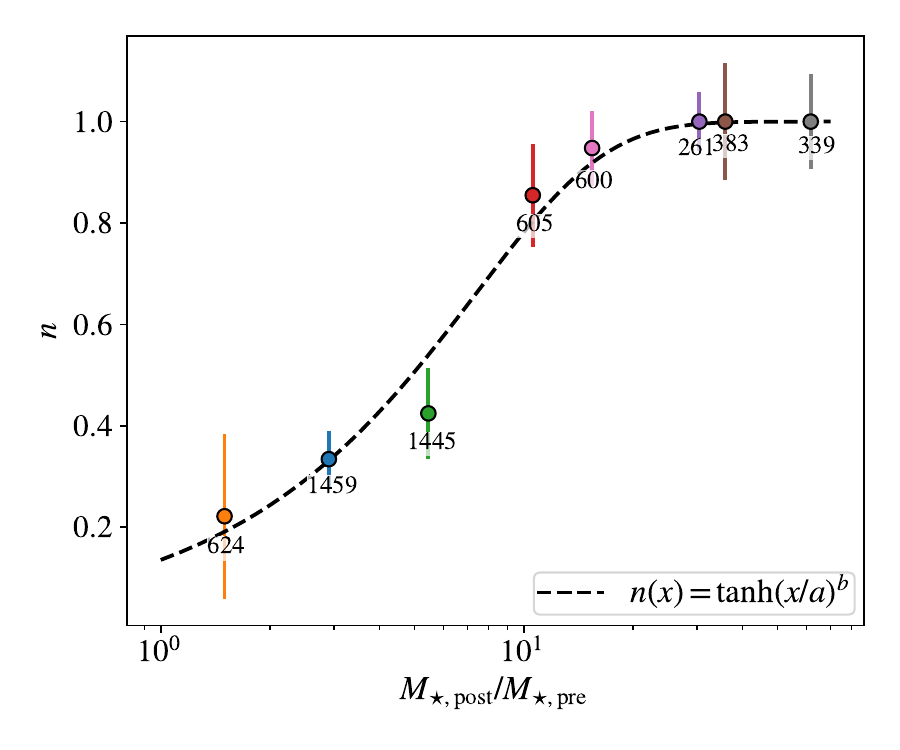}}
     \caption{The values of the coreNFW inner slope parameter $n$ found for the \textsc{edge}2 haloes as a function of the $\mathrm{M}_{\star,\mathrm{post}} / \mathrm{M}_{\star,\mathrm{pre}}$ ratios. A value of $n=0$ means an NFW cusp, while a value of $n=1$ is a perfect core. The error bars show the 1$\sigma$ errors on the parameter. The dashed line shows the fitting function given by Eq.~\eqref{eqn}.} 
     \label{n_function_fit}
\end{figure}

Fig.~\ref{n_function_fit} shows the values of $n$ found for each \textsc{edge}2 halo as a function of $\mathrm{M}_{\star, \mathrm{post}}/\mathrm{M}_{\star, \mathrm{pre}}$. The values of $n$ increase with $\mathrm{M}_{\star, \mathrm{post}}/\mathrm{M}_{\star, \mathrm{pre}}$, capturing the tight relation between our ratio and the amount of `coreness' in the haloes' inner regions. 
To parameterise the changes in $n$, we adopt an analytical function of the form
\begin{equation} \label{eqn}
    n(x) = \mathrm{tanh}(x/a)^b,
\end{equation} 
where $x=\mathrm{M}_{\star, \mathrm{post}}/\mathrm{M}_{\star, \mathrm{pre}}$. Fitting to the inferred values of $n$ for the halos gives $a=10.27 \pm 1.54$ and $b=0.86 \pm 0.13$, valid within the range covered by our simulation data ($1 \lesssim \mathrm{M}_{\star, \mathrm{post}}/\mathrm{M}_{\star, \mathrm{pre}} \lesssim 70$, and $\mathrm{M}_{\star}<1.5\times 10^7\,M_{\odot}$).

\section{Conclusion and Discussion} \label{sec:discussion}

We presented evidence from the \textsc{edge} suite of hydrodynamical simulations that the extent of baryon-driven dark matter core creation 
strongly depends on the timing of star formation relative to reionisation. We focused our study on $\rho_{\mathrm{DM}}(150\,\mathrm{pc})$, the $z=0$ density of dark matter at $150\,\mathrm{pc}$ distance from the halo centre, motivated by the existence of observational estimates in dwarf irregulars and spheroidals at this radius. In the following discussion we refer to this density as $\rho_{\mathrm{DM}}$, leaving the radius implicit.  We find the \textsc{edge} stellar-to-halo mass ratio ($\textrm{M}_{\star} / \textrm{M}_{200}$) correlates with $\rho_{\mathrm{DM}}$ in approximately the way seen in observations. However, we also find that the simulated density correlates much more tightly with a new quantity, $\mathrm{M}_{\star,\mathrm{post}} / \mathrm{M}_{\star,\mathrm{pre}}$, i.e. the ratio between post- and pre-reionisation stellar mass, showing that the timing of star formation is a crucial consideration.

In addition to existing \textsc{edge}1 simulations \citep{Agertz2020EDGE,Rey2020,Orkney_2021}, we introduced a new generation of \textsc{edge}2 simulations which extend to higher mass dwarfs, and include non-equilibrium cooling. Furthermore, all \textsc{edge}2 simulations include photo-ionization feedback using radiative transfer, whereas the majority of simulations in the  \textsc{edge}1 suite (including all those presented here) were performed without such a feedback channel. The correlation between $\rho_{\mathrm{DM}}$ and $\mathrm{M}_{\star,\mathrm{post}} / \mathrm{M}_{\star,\mathrm{pre}}$ remains tight when including both suites of simulations, despite radiative transfer and non-equilibrium cooling making a major difference to other observable properties \citep{Agertz2020EDGE}. This gives us some confidence that our key conclusions are robust against changes in feedback details. Nonetheless it will clearly be of considerable interest to investigate our results using alternative state-of-the-art codes \citep[e.g.][]{Jeon_2017, Wheeler_2019, Munshi_2021, Gutcke_2022}. All \textsc{edge} dwarfs considered are in broad agreement with the stellar-mass halo-mass relation inferred from observations, albeit in a mass range where there is considerable uncertainty.

As discussed in the Introduction, previous works have focussed on $\textrm{M}_{\star}/\textrm{M}_{200}$ as the primary determinant of core creation. The numerator of this ratio focuses on the total available feedback energy (roughly proportional to $\mathrm{M}_{\star}$). By contrast, the numerator of our revised ratio $\mathrm{M}_{\star,\mathrm{post}} $ is a proxy for {\it sustained} star formation, based on the assumption that dwarfs surviving reionisation will continue to form stars over an extended period. Forming stars in a  concentrated burst at early times is not an effective way to soften the dark matter cusp \citep{Gnedin_2002, Pontzen_2012, Chan_2015, Jackson_2024}; rather, sustained cycles of star formation, gas expulsion and collapse are required to produce a long-term fluctuating gravitational potential. 

$\textrm{M}_{200}$, which is in the denominator of the traditional ratio, is relevant because it is proportional to the depth of the gravitational potential in which the feedback operates. However, the actual potential to be overcome is not that of the whole halo but rather the concentrated central cusp (i.e. particles do not need to be expelled from the halo as a whole). This central concentration depends on the early accretion history \citep{Ludlow_2014}, for which we use $\mathrm{M}_{\star,\mathrm{pre}}$ as a proxy that is in principle observable (we discuss observational prospects further below). We also explored the use of other proxies for the central density, such as the maximum circular velocity ($v_{\mathrm{max}}$) \citep{Prada_2012}, but found that $\mathrm{M}_{\star,\mathrm{post}}/\mathrm{M}_{\star,\mathrm{pre}}$ correlates most tightly with $\rho_{\mathrm{DM}}$.

Our stellar masses are calculated across all progenitor branches, and hence $\mathrm{M}_{\star,\mathrm{pre}}$ is sensitive to the early collapse of the region as a whole rather than solely of the major progenitor. 
The correlation with $\rho_{\mathrm{DM}}$ becomes less tight when $\mathrm{M}_{\star,\mathrm{pre}}$ is evaluated using only the stars formed on the main progenitor branch. A possible reason for this sensitivity to the history of the entire progenitor tree, rather than solely the major progenitor, is that mergers with a steeply cusped halo can increase the central density of the galaxy \citep{Laporte_2015, Onorbe_2015, Orkney_2021}. We hypothesize that, by incorporating  information about all the progenitors in $\mathrm{M}_{\star, \mathrm{post}}/\mathrm{M}_{\star, \mathrm{pre}}$, this effect is implicitly accounted for. 

More broadly, the fact that $\mathrm{M}_{\star,\mathrm{post}}/\mathrm{M}_{\star,\mathrm{pre}}$ is a good predictor of dark matter density at $z=0$ reflects the importance of the interplay between reionisation and star formation in low-mass galaxies. In future work, we will explore other quantities which could connect to the halo central density and perform comprehensive studies of the role of formation history by employing genetically modified galaxies (such as late / early formers). 
We expect $\mathrm{M}_{\star,\mathrm{post}}/\mathrm{M}_{\star,\mathrm{pre}}$ to play a less essential role in higher mass galaxies, e.g. in determining the cuspiness of Milky-Way mass haloes where stellar feedback is likely ineffective in core creation for different physical reasons \citep[e.g.][]{Di_Cintio_2013, Chan_2015, Tollet_2016}.

It would be challenging, but very well-motivated, to estimate observationally a quantity like $\mathrm{M}_{\star, \mathrm{pre}}$. The requirement for $\lesssim$Gyr resolution in a star formation history reconstruction over the Hubble time constitutes the major challenge. Even now, however, one can make a qualitative comparison with observed galaxies given recent progress in reconstructing dwarf galaxy star formation histories \citep[e.g.][]{Cole_2014, Read_2019, McQuinn_2023}. 
From Fig.~5 in \citet{Read_2019} and Fig. 11 in \citet{Bouche_2022}, one may indeed infer that galaxies whose star formation stopped a long time ago typically have steeper cusps, while dwarfs with more extended star-formation have shallower cores. For instance, Aquarius and LeoII have similar stellar masses and halo masses, but Aquarius has a more extended star formation history and a lower central density, as expected if sustained star formation at late times is necessary for the formation of cores. 

On the other hand, the observational picture is as yet far from clear-cut; for example, LeoI \citep{Read_2019} seems to have a cuspy profile despite having formed stars until relatively recently. One possibility is that LeoI underwent a recent merger \citep{Ruiz-Lara_2021}, enabling it to rebuild a dark matter density cusp \citep{Laporte_2015, Orkney_2021}. This reinforces the need, mentioned above, to explore a greater variety of simulated histories in future work.
There has also been speculation about the role that a black hole at the centre of LeoI may have played in its evolution \citep[e.g.][]{Bustamante_Rosell_2021} (although see also \citep{Pascale_2024}), and we are in the process of incorporating black holes into the \textsc{edge} code for future work. 

As discussed in Section \ref{sec:intro}, the central dark matter densities of low-mass dwarf galaxies hold enormous potential to shed light on particle physics. As such, understanding the confounding effects of baryonic feedback is an essential project, in which the timing of star formation now appears a critical component. To incorporate this rigorously, it will be necessary to interpret forthcoming observations in the light of a broad array of simulations covering a wide variety of formation histories, and making use of codes with a range of alternative feedback prescriptions. However the tight correlations seen in the present work give considerable hope that, in spite of the complexities, baryonic core creation is an essentially predictable process.

\section*{Acknowledgements}
CM would like to thank Dr Jason L. Sanders for the helpful feedback and discussions. CM is supported by the Science and Technology Facilities Council. JIR would like to acknowledge support from STFC grants ST/Y002865/1 and ST/Y002857/1.
OA acknowledges support from the Knut and Alice Wallenberg Foundation, the Swedish Research Council (grant 2019-04659), and the Swedish National Space Agency (SNSA Dnr 2023-00164). 
MR is supported by the Beecroft Fellowship funded by Adrian Beecroft.
ET acknowledges the UKRI Science and Technology Facilities Council (STFC) for support (grant ST/V50712X/1).
This study was supported by the European Research Council (ERC) under the European Union’s Horizon 2020 research and innovation programme (grant agreement No. 818085 GMGalaxies). This work was performed in part using the DiRAC Data Intensive service at Leicester, operated by the University of Leicester IT Services, which forms part of the STFC DiRAC HPC Facility (www.dirac.ac.uk). The equipment was funded by BEIS capital funding via STFC capital grants ST/K000373/1 and ST/R002363/1 and STFC DiRAC Operations grant ST/R001014/1. DiRAC is part of the National e-Infrastructure. This work was partially enabled by funding from the UCL Cosmoparticle Initiative.

\section*{Author Contributions}
The contributions from the authors are listed below using key-words based on the CRediT (Contribution Roles Taxonomy) system.
\textbf{CM:} conceptualization; investigation; methodology; software; formal analysis; visualisation; writing - original draft, review $\&$ editing.
\textbf{AP:} conceptualization; methodology; validation and interpretation; supervision; resources; writing - original draft, review $\&$ editing; funding acquisition.
\textbf{JIR:} conceptualization; methodology; validation and interpretation; writing - review $\&$ editing.
\textbf{OA:} methodology; writing - review $\&$ editing.
\textbf{MPR:} software, data curation, resources.
\textbf{ET:} writing - review $\&$ editing.
\textbf{SYK:} writing - review $\&$ editing.
\textbf{EIG:} data curation; writing - review $\&$ editing.

\section*{Data Availability}
Data may be shared upon reasonable request.



\bibliographystyle{mnras}
\bibliography{example} 

\begin{thebibliography}{}
\makeatletter
\relax
\def\mn@urlcharsother{\let\do\@makeother \do\$\do\&\do\#\do\^\do\_\do\%\do\~}
\def\mn@doi{\begingroup\mn@urlcharsother \@ifnextchar [ {\mn@doi@} {\mn@doi@[]}}
\def\mn@doi@[#1]#2{\def\@tempa{#1}\ifx\@tempa\@empty \href {http://dx.doi.org/#2} {doi:#2}\else \href {http://dx.doi.org/#2} {#1}\fi \endgroup}
\def\mn@eprint#1#2{\mn@eprint@#1:#2::\@nil}
\def\mn@eprint@arXiv#1{\href {http://arxiv.org/abs/#1} {{\tt arXiv:#1}}}
\def\mn@eprint@dblp#1{\href {http://dblp.uni-trier.de/rec/bibtex/#1.xml} {dblp:#1}}
\def\mn@eprint@#1:#2:#3:#4\@nil{\def\@tempa {#1}\def\@tempb {#2}\def\@tempc {#3}\ifx \@tempc \@empty \let \@tempc \@tempb \let \@tempb \@tempa \fi \ifx \@tempb \@empty \def\@tempb {arXiv}\fi \@ifundefined {mn@eprint@\@tempb}{\@tempb:\@tempc}{\expandafter \expandafter \csname mn@eprint@\@tempb\endcsname \expandafter{\@tempc}}}

\bibitem[\protect\citeauthoryear{Adams et~al.,}{Adams et~al.}{2014}]{Adams_2014}
Adams J.~J.,  et~al., 2014, \mn@doi [The Astrophysical Journal] {10.1088/0004-637x/789/1/63}, 789, 63

\bibitem[\protect\citeauthoryear{Agertz et~al.,}{Agertz et~al.}{2020}]{Agertz2020EDGE}
Agertz O.,  et~al., 2020, \mn@doi [MNRAS] {10.1093/mnras/stz3053}, 491, 1656

\bibitem[\protect\citeauthoryear{Aubert, Pichon  \& Colombi}{Aubert et~al.}{2004}]{Aubert2004}
Aubert D.,  Pichon C.,   Colombi S.,  2004, \mn@doi [MNRAS] {10.1111/j.1365-2966.2004.07883.x}, 352, 376

\bibitem[\protect\citeauthoryear{Behroozi, Conroy  \& Wechsler}{Behroozi et~al.}{2010}]{Behroozi_2010}
Behroozi P.~S.,  Conroy C.,   Wechsler R.~H.,  2010, \mn@doi [The Astrophysical Journal] {10.1088/0004-637x/717/1/379}, 717, 379–403

\bibitem[\protect\citeauthoryear{Benson, Lacey, Baugh, Cole  \& Frenk}{Benson et~al.}{2002}]{Benson_2002}
Benson A.~J.,  Lacey C.~G.,  Baugh C.~M.,  Cole S.,   Frenk C.~S.,  2002, \mn@doi [MNRAS] {10.1046/j.1365-8711.2002.05387.x}, 333, 156–176

\bibitem[\protect\citeauthoryear{Benítez-Llambay, Frenk, Ludlow  \& Navarro}{Benítez-Llambay et~al.}{2019}]{Benitez_Llambay_2019}
Benítez-Llambay A.,  Frenk C.~S.,  Ludlow A.~D.,   Navarro J.~F.,  2019, \mn@doi [MNRAS] {10.1093/mnras/stz1890}, 488, 2387–2404

\bibitem[\protect\citeauthoryear{Binney \& Mamon}{Binney \& Mamon}{1982}]{Binney_1982}
Binney J.,  Mamon G.~A.,  1982, \mn@doi [MNRAS] {10.1093/mnras/200.2.361}, 200, 361

\bibitem[\protect\citeauthoryear{Bouché et~al.,}{Bouché et~al.}{2022}]{Bouche_2022}
Bouché N.~F.,  et~al., 2022, \mn@doi [A&A] {10.1051/0004-6361/202141762}, 658, A76

\bibitem[\protect\citeauthoryear{Bullock \& Boylan-Kolchin}{Bullock \& Boylan-Kolchin}{2017}]{Bullock_2017}
Bullock J.~S.,  Boylan-Kolchin M.,  2017, \mn@doi [Annual Review of Astronomy and Astrophysics] {10.1146/annurev-astro-091916-055313}, 55, 343–387

\bibitem[\protect\citeauthoryear{Bustamante-Rosell, Noyola, Gebhardt, Fabricius, Mazzalay, Thomas  \& Zeimann}{Bustamante-Rosell et~al.}{2021}]{Bustamante_Rosell_2021}
Bustamante-Rosell M.~J.,  Noyola E.,  Gebhardt K.,  Fabricius M.~H.,  Mazzalay X.,  Thomas J.,   Zeimann G.,  2021, \mn@doi [The Astrophysical Journal] {10.3847/1538-4357/ac0c79}, 921, 107

\bibitem[\protect\citeauthoryear{Chan, Kereš, Oñorbe, Hopkins, Muratov, Faucher-Giguère  \& Quataert}{Chan et~al.}{2015}]{Chan_2015}
Chan T.~K.,  Kereš D.,  Oñorbe J.,  Hopkins P.~F.,  Muratov A.~L.,  Faucher-Giguère C.-A.,   Quataert E.,  2015, \mn@doi [MNRAS] {10.1093/mnras/stv2165}, 454, 2981–3001

\bibitem[\protect\citeauthoryear{{Christensen}, {Brooks}, {Munshi}, {Riggs}, {Van Nest}, {Akins}, {Quinn}  \& {Chamberland}}{{Christensen} et~al.}{2024}]{Christensen_2024}
{Christensen} C.~R.,  {Brooks} A.~M.,  {Munshi} F.,  {Riggs} C.,  {Van Nest} J.,  {Akins} H.,  {Quinn} T.~R.,   {Chamberland} L.,  2024, \mn@doi [\apj] {10.3847/1538-4357/ad0c5a}, \href {https://ui.adsabs.harvard.edu/abs/2024ApJ...961..236C} {961, 236}

\bibitem[\protect\citeauthoryear{Cole, Weisz, Dolphin, Skillman, McConnachie, Brooks  \& Leaman}{Cole et~al.}{2014}]{Cole_2014}
Cole A.~A.,  Weisz D.~R.,  Dolphin A.~E.,  Skillman E.~D.,  McConnachie A.~W.,  Brooks A.~M.,   Leaman R.,  2014, \mn@doi [The Astrophysical Journal] {10.1088/0004-637x/795/1/54}, 795, 54

\bibitem[\protect\citeauthoryear{Collins \& Read}{Collins \& Read}{2022}]{Collins_2022}
Collins M. L.~M.,  Read J.~I.,  2022, \mn@doi [Nature Astronomy] {10.1038/s41550-022-01657-4}, 6, 647–658

\bibitem[\protect\citeauthoryear{Cooke et~al.,}{Cooke et~al.}{2022}]{Cooke_2022}
Cooke L.~H.,  et~al., 2022, \mn@doi [MNRAS] {10.1093/mnras/stac588}, 512, 1012–1031

\bibitem[\protect\citeauthoryear{Courty \& Alimi}{Courty \& Alimi}{2004}]{Courty2004}
Courty S.,  Alimi J.~M.,  2004, \mn@doi [A\&A] {10.1051/0004-6361:20031736}, 416, 875

\bibitem[\protect\citeauthoryear{{Danieli}, {Greene}, {Carlsten}, {Jiang}, {Beaton}  \& {Goulding}}{{Danieli} et~al.}{2023}]{Danieli_2023}
{Danieli} S.,  {Greene} J.~E.,  {Carlsten} S.,  {Jiang} F.,  {Beaton} R.,   {Goulding} A.~D.,  2023, \mn@doi [\apj] {10.3847/1538-4357/acefbd}, \href {https://ui.adsabs.harvard.edu/abs/2023ApJ...956....6D} {956, 6}

\bibitem[\protect\citeauthoryear{De~Leo, Read, Noel, Erkal, Massana  \& Carrera}{De~Leo et~al.}{2023}]{Deleo_2023}
De~Leo M.,  Read J.~I.,  Noel N. E.~D.,  Erkal D.,  Massana P.,   Carrera R.,  2023, Surviving the Waves: evidence for a Dark Matter cusp in the tidally disrupting Small Magellanic Cloud (\mn@eprint {arXiv} {2303.08838})

\bibitem[\protect\citeauthoryear{{Dekel} \& {Silk}}{{Dekel} \& {Silk}}{1986}]{Dekel_1986}
{Dekel} A.,  {Silk} J.,  1986, \mn@doi [\apj] {10.1086/164050}, \href {https://ui.adsabs.harvard.edu/abs/1986ApJ...303...39D} {303, 39}

\bibitem[\protect\citeauthoryear{Di~Cintio, Brook, Macciò, Stinson, Knebe, Dutton  \& Wadsley}{Di~Cintio et~al.}{2013}]{Di_Cintio_2013}
Di~Cintio A.,  Brook C.~B.,  Macciò A.~V.,  Stinson G.~S.,  Knebe A.,  Dutton A.~A.,   Wadsley J.,  2013, \mn@doi [MNRAS] {10.1093/mnras/stt1891}, 437, 415–423

\bibitem[\protect\citeauthoryear{{Di Cintio}, {Brook}, {Macci{\`o}}, {Stinson}, {Knebe}, {Dutton}  \& {Wadsley}}{{Di Cintio} et~al.}{2014}]{DiCintio14}
{Di Cintio} A.,  {Brook} C.~B.,  {Macci{\`o}} A.~V.,  {Stinson} G.~S.,  {Knebe} A.,  {Dutton} A.~A.,   {Wadsley} J.,  2014, \mn@doi [\mnras] {10.1093/mnras/stt1891}, \href {https://ui.adsabs.harvard.edu/abs/2014MNRAS.437..415D} {437, 415}

\bibitem[\protect\citeauthoryear{Dutton, Macciò, Buck, Dixon, Blank  \& Obreja}{Dutton et~al.}{2019}]{Dutton_2019}
Dutton A.~A.,  Macciò A.~V.,  Buck T.,  Dixon K.~L.,  Blank M.,   Obreja A.,  2019, \mn@doi [MNRAS] {10.1093/mnras/stz889}, 486, 655–671

\bibitem[\protect\citeauthoryear{Efstathiou}{Efstathiou}{1992}]{Efstathiou_1992}
Efstathiou G.,  1992, \mn@doi [MNRAS] {10.1093/mnras/256.1.43P}, 256, 43P

\bibitem[\protect\citeauthoryear{Efstathiou}{Efstathiou}{2000}]{Efstathiou_2000}
Efstathiou G.,  2000, \mn@doi [MNRAS] {10.1046/j.1365-8711.2000.03665.x}, 317, 697–719

\bibitem[\protect\citeauthoryear{Eisenstein \& Hut}{Eisenstein \& Hut}{1998}]{Eisenstein1998}
Eisenstein D.~J.,  Hut P.,  1998, \mn@doi [ApJ] {10.1086/305535}, 498, 137

\bibitem[\protect\citeauthoryear{El‐Zant, Shlosman  \& Hoffman}{El‐Zant et~al.}{2001}]{El_Zant_2001}
El‐Zant A.,  Shlosman I.,   Hoffman Y.,  2001, \mn@doi [The Astrophysical Journal] {10.1086/322516}, 560, 636–643

\bibitem[\protect\citeauthoryear{Emami, Siana, Weisz, Johnson, Ma  \& El-Badry}{Emami et~al.}{2019}]{Emami_2019}
Emami N.,  Siana B.,  Weisz D.~R.,  Johnson B.~D.,  Ma X.,   El-Badry K.,  2019, \mn@doi [The Astrophysical Journal] {10.3847/1538-4357/ab211a}, 881, 71

\bibitem[\protect\citeauthoryear{Fattahi, Navarro, Frenk, Oman, Sawala  \& Schaller}{Fattahi et~al.}{2018}]{Fattahi_2018}
Fattahi A.,  Navarro J.~F.,  Frenk C.~S.,  Oman K.~A.,  Sawala T.,   Schaller M.,  2018, \mn@doi [MNRAS] {10.1093/mnras/sty408}, 476, 3816–3836

\bibitem[\protect\citeauthoryear{{Faucher-Gigu{\`e}re}}{{Faucher-Gigu{\`e}re}}{2020}]{Faucher-Giguere2020}
{Faucher-Gigu{\`e}re} C.-A.,  2020, \mn@doi [MNRAS] {10.1093/mnras/staa302}, 493, 1614

\bibitem[\protect\citeauthoryear{{Garrison-Kimmel}, {Bullock}, {Boylan-Kolchin}  \& {Bardwell}}{{Garrison-Kimmel} et~al.}{2017}]{Garrison-Kimmel_2017}
{Garrison-Kimmel} S.,  {Bullock} J.~S.,  {Boylan-Kolchin} M.,   {Bardwell} E.,  2017, \mn@doi [\mnras] {10.1093/mnras/stw2564}, \href {https://ui.adsabs.harvard.edu/abs/2017MNRAS.464.3108G} {464, 3108}

\bibitem[\protect\citeauthoryear{Gnedin \& Zhao}{Gnedin \& Zhao}{2002}]{Gnedin_2002}
Gnedin O.~Y.,  Zhao H.,  2002, \mn@doi [MNRAS] {10.1046/j.1365-8711.2002.05361.x}, 333, 299–306

\bibitem[\protect\citeauthoryear{Governato et~al.,}{Governato et~al.}{2010}]{Governato_2010}
Governato F.,  et~al., 2010, \mn@doi [Nature] {10.1038/nature08640}, 463, 203–206

\bibitem[\protect\citeauthoryear{Gray et~al.,}{Gray et~al.}{2024}]{gray_2024}
Gray E.~I.,  et~al., 2024, EDGE: A new model for Nuclear Star Cluster formation in dwarf galaxies (\mn@eprint {arXiv} {2405.19286}), \url {https://arxiv.org/abs/2405.19286}

\bibitem[\protect\citeauthoryear{Gutcke, Pakmor, Naab  \& Springel}{Gutcke et~al.}{2022}]{Gutcke_2022}
Gutcke T.~A.,  Pakmor R.,  Naab T.,   Springel V.,  2022, \mn@doi [MNRAS] {10.1093/mnras/stac867}, 513, 1372–1385

\bibitem[\protect\citeauthoryear{Haardt \& Madau}{Haardt \& Madau}{1996}]{Haardt1996}
Haardt F.,  Madau P.,  1996, \mn@doi [ApJ] {10.1086/177035}, 461, 20

\bibitem[\protect\citeauthoryear{{Hu}, {Barkana}  \& {Gruzinov}}{{Hu} et~al.}{2000}]{Hu00}
{Hu} W.,  {Barkana} R.,   {Gruzinov} A.,  2000, \mn@doi [\prl] {10.1103/PhysRevLett.85.1158}, \href {https://ui.adsabs.harvard.edu/abs/2000PhRvL..85.1158H} {85, 1158}

\bibitem[\protect\citeauthoryear{{Jackson} et~al.,}{{Jackson} et~al.}{2024}]{Jackson_2024}
{Jackson} R.~A.,  et~al., 2024, \mn@doi [\mnras] {10.1093/mnras/stae056}, \href {https://ui.adsabs.harvard.edu/abs/2024MNRAS.528.1655J} {528, 1655}

\bibitem[\protect\citeauthoryear{Jeon, Besla  \& Bromm}{Jeon et~al.}{2017}]{Jeon_2017}
Jeon M.,  Besla G.,   Bromm V.,  2017, \mn@doi [The Astrophysical Journal] {10.3847/1538-4357/aa8c80}, 848, 85

\bibitem[\protect\citeauthoryear{Kauffmann}{Kauffmann}{2014}]{Kauffmann_2014}
Kauffmann G.,  2014, MNRAS, 441, 2717

\bibitem[\protect\citeauthoryear{{Kim} et~al.,}{{Kim} et~al.}{2024}]{Kim_2024}
{Kim} S.~Y.,  et~al., 2024, \mn@doi [arXiv e-prints] {10.48550/arXiv.2408.15214}, \href {https://ui.adsabs.harvard.edu/abs/2024arXiv240815214K} {p. arXiv:2408.15214}

\bibitem[\protect\citeauthoryear{{Laporte} \& {Penarrubia}}{{Laporte} \& {Penarrubia}}{2015}]{Laporte_2015}
{Laporte} C.~F.~P.,  {Penarrubia} J.,  2015, \mn@doi [\mnras] {10.1093/mnrasl/slv008}, \href {https://ui.adsabs.harvard.edu/abs/2015MNRAS.449L..90L} {449, L90}

\bibitem[\protect\citeauthoryear{Ludlow, Navarro, Angulo, Boylan-Kolchin, Springel, Frenk  \& White}{Ludlow et~al.}{2014}]{Ludlow_2014}
Ludlow A.~D.,  Navarro J.~F.,  Angulo R.~E.,  Boylan-Kolchin M.,  Springel V.,  Frenk C.,   White S. D.~M.,  2014, \mn@doi [MNRAS] {10.1093/mnras/stu483}, 441, 378–388

\bibitem[\protect\citeauthoryear{Marasco, Oman, Navarro, Frenk  \& Oosterloo}{Marasco et~al.}{2018}]{Marasco_2018}
Marasco A.,  Oman K.~A.,  Navarro J.~F.,  Frenk C.~S.,   Oosterloo T.,  2018, \mn@doi [MNRAS] {10.1093/mnras/sty354}, 476, 2168–2176

\bibitem[\protect\citeauthoryear{Mashchenko, Wadsley  \& Couchman}{Mashchenko et~al.}{2008}]{Mashchenko_2008}
Mashchenko S.,  Wadsley J.,   Couchman H. M.~P.,  2008, \mn@doi [Science] {10.1126/science.1148666}, 319, 174–177

\bibitem[\protect\citeauthoryear{McQuinn, Lelli, Skillman, Dolphin, McGaugh  \& Williams}{McQuinn et~al.}{2015}]{McQuinn_2015}
McQuinn K. B.~W.,  Lelli F.,  Skillman E.~D.,  Dolphin A.~E.,  McGaugh S.~S.,   Williams B.~F.,  2015, \mn@doi [MNRAS] {10.1093/mnras/stv841}, 450, 3886

\bibitem[\protect\citeauthoryear{McQuinn, Mao, Tollerud, Cohen, Shih, Buckley  \& Dolphin}{McQuinn et~al.}{2024}]{McQuinn_2023}
McQuinn K. B.~W.,  Mao Y.-Y.,  Tollerud E.~J.,  Cohen R.~E.,  Shih D.,  Buckley M.~R.,   Dolphin A.~E.,  2024, \mn@doi [Astrophys. J.] {10.3847/1538-4357/ad429b}, 967, 161

\bibitem[\protect\citeauthoryear{Mina, Mota  \& Winther}{Mina et~al.}{2022}]{Mina_2022}
Mina M.,  Mota D.~F.,   Winther H.~A.,  2022, \mn@doi [A&A] {10.1051/0004-6361/202038876}, 662, A29

\bibitem[\protect\citeauthoryear{{Moster}, {Naab}  \& {White}}{{Moster} et~al.}{2013}]{Moster_2013}
{Moster} B.~P.,  {Naab} T.,   {White} S. D.~M.,  2013, \mn@doi [\mnras] {10.1093/mnras/sts261}, \href {https://ui.adsabs.harvard.edu/abs/2013MNRAS.428.3121M} {428, 3121}

\bibitem[\protect\citeauthoryear{Muni, Pontzen, Sanders, Rey, Read  \& Agertz}{Muni et~al.}{2023}]{Muni_2023}
Muni C.,  Pontzen A.,  Sanders J.~L.,  Rey M.~P.,  Read J.~I.,   Agertz O.,  2023, \mn@doi [MNRAS] {10.1093/mnras/stad3835}, 527, 9250–9262

\bibitem[\protect\citeauthoryear{{Munshi}, {Brooks}, {Applebaum}, {Christensen}, {Quinn}  \& {Sligh}}{{Munshi} et~al.}{2021}]{Munshi_2021}
{Munshi} F.,  {Brooks} A.~M.,  {Applebaum} E.,  {Christensen} C.~R.,  {Quinn} T.,   {Sligh} S.,  2021, \mn@doi [\apj] {10.3847/1538-4357/ac0db6}, \href {https://ui.adsabs.harvard.edu/abs/2021ApJ...923...35M} {923, 35}

\bibitem[\protect\citeauthoryear{Mutlu-Pakdil et~al.,}{Mutlu-Pakdil et~al.}{2019}]{Mutlu_Pakdil_2019}
Mutlu-Pakdil B.,  et~al., 2019, \mn@doi [The Astrophysical Journal] {10.3847/1538-4357/ab45ec}, 885, 53

\bibitem[\protect\citeauthoryear{Nadler et~al.,}{Nadler et~al.}{2020}]{Nadler_2020}
Nadler E.~O.,  et~al., 2020, \mn@doi [The Astrophysical Journal] {10.3847/1538-4357/ab846a}, 893, 48

\bibitem[\protect\citeauthoryear{Navarro, Eke  \& Frenk}{Navarro et~al.}{1996}]{Navarro_1996}
Navarro J.~F.,  Eke V.~R.,   Frenk C.~S.,  1996, \mn@doi [MNRAS] {10.1093/mnras/283.3.l72}, 283, L72–L78

\bibitem[\protect\citeauthoryear{Navarro, Frenk  \& White}{Navarro et~al.}{1997}]{Navarro_1997}
Navarro J.~F.,  Frenk C.~S.,   White S. D.~M.,  1997, \mn@doi [The Astrophysical Journal] {10.1086/304888}, 490, 493–508

\bibitem[\protect\citeauthoryear{Newton et~al.,}{Newton et~al.}{2021}]{Newton_2021}
Newton O.,  et~al., 2021, \mn@doi [Journal of Cosmology and Astroparticle Physics] {10.1088/1475-7516/2021/08/062}, 2021, 062

\bibitem[\protect\citeauthoryear{Nipoti \& Binney}{Nipoti \& Binney}{2014}]{Nipoti_2014}
Nipoti C.,  Binney J.,  2014, \mn@doi [MNRAS] {10.1093/mnras/stu2217}, 446, 1820–1828

\bibitem[\protect\citeauthoryear{Oman et~al.,}{Oman et~al.}{2015}]{Oman_2015}
Oman K.~A.,  et~al., 2015, \mn@doi [MNRAS] {10.1093/mnras/stv1504}, 452, 3650–3665

\bibitem[\protect\citeauthoryear{Oman, Marasco, Navarro, Frenk, Schaye  \& Benítez-Llambay}{Oman et~al.}{2018}]{Oman_2018}
Oman K.~A.,  Marasco A.,  Navarro J.~F.,  Frenk C.~S.,  Schaye J.,   Benítez-Llambay A.,  2018, \mn@doi [MNRAS] {10.1093/mnras/sty2687}, 482, 821–847

\bibitem[\protect\citeauthoryear{Orkney et~al.,}{Orkney et~al.}{2021}]{Orkney_2021}
Orkney M. D.~A.,  et~al., 2021, \mn@doi [MNRAS] {10.1093/mnras/stab1066}, 504, 3509–3522

\bibitem[\protect\citeauthoryear{Oñorbe, Boylan-Kolchin, Bullock, Hopkins, Kereš, Faucher-Giguère, Quataert  \& Murray}{Oñorbe et~al.}{2015}]{Onorbe_2015}
Oñorbe J.,  Boylan-Kolchin M.,  Bullock J.~S.,  Hopkins P.~F.,  Kereš D.,  Faucher-Giguère C.-A.,  Quataert E.,   Murray N.,  2015, \mn@doi [MNRAS] {10.1093/mnras/stv2072}, 454, 2092–2106

\bibitem[\protect\citeauthoryear{Pace, Erkal  \& Li}{Pace et~al.}{2022}]{Pace_2022}
Pace A.~B.,  Erkal D.,   Li T.~S.,  2022, \mn@doi [The Astrophysical Journal] {10.3847/1538-4357/ac997b}, 940, 136

\bibitem[\protect\citeauthoryear{Pascale, Nipoti, Calura  \& Della~Croce}{Pascale et~al.}{2024}]{Pascale_2024}
Pascale R.,  Nipoti C.,  Calura F.,   Della~Croce A.,  2024, \mn@doi [Astronomy &amp; Astrophysics] {10.1051/0004-6361/202449620}, 684, L19

\bibitem[\protect\citeauthoryear{Peñarrubia, Pontzen, Walker  \& Koposov}{Peñarrubia et~al.}{2012}]{Penarrubia_2012}
Peñarrubia J.,  Pontzen A.,  Walker M.~G.,   Koposov S.~E.,  2012, \mn@doi [The Astrophysical Journal] {10.1088/2041-8205/759/2/l42}, 759, L42

\bibitem[\protect\citeauthoryear{{Planck Collaboration} et~al.,}{{Planck Collaboration} et~al.}{2014}]{PlanckCollaboration2014}
{Planck Collaboration} et~al., 2014, \mn@doi [A\&A] {10.1051/0004-6361/201321591}, 571, A16

\bibitem[\protect\citeauthoryear{Pontzen \& Governato}{Pontzen \& Governato}{2012}]{Pontzen_2012}
Pontzen A.,  Governato F.,  2012, \mn@doi [MNRAS] {10.1111/j.1365-2966.2012.20571.x}, 421, 3464–3471

\bibitem[\protect\citeauthoryear{Pontzen \& Governato}{Pontzen \& Governato}{2014}]{Pontzen_2014}
Pontzen A.,  Governato F.,  2014, \mn@doi [Nature] {10.1038/nature12953}, 506, 171–178

\bibitem[\protect\citeauthoryear{Pontzen \& Tremmel}{Pontzen \& Tremmel}{2018}]{Pontzen2018}
Pontzen A.,  Tremmel M.,  2018, \mn@doi [ApJS] {10.3847/1538-4365/aac832}, 237, 23

\bibitem[\protect\citeauthoryear{{Pontzen}, {Ro{\v{s}}kar}, {Stinson}  \& {Woods}}{{Pontzen} et~al.}{2013}]{Pontzen2013}
{Pontzen} A.,  {Ro{\v{s}}kar} R.,  {Stinson} G.,   {Woods} R.,  2013, {pynbody: N-Body/SPH analysis for python}, Astrophysics Source Code Library, record ascl:1305.002

\bibitem[\protect\citeauthoryear{Power, Navarro, Jenkins, Frenk, White, Springel, Stadel  \& Quinn}{Power et~al.}{2003}]{Power2003}
Power C.,  Navarro J.~F.,  Jenkins A.,  Frenk C.~S.,  White S. D.~M.,  Springel V.,  Stadel J.,   Quinn T.,  2003, \mn@doi [MNRAS] {10.1046/j.1365-8711.2003.05925.x}, 338, 14

\bibitem[\protect\citeauthoryear{Prada, Klypin, Cuesta, Betancort-Rijo  \& Primack}{Prada et~al.}{2012}]{Prada_2012}
Prada F.,  Klypin A.~A.,  Cuesta A.~J.,  Betancort-Rijo J.~E.,   Primack J.,  2012, \mn@doi [MNRAS] {10.1111/j.1365-2966.2012.21007.x}, 423, 3018–3030

\bibitem[\protect\citeauthoryear{Read \& Gilmore}{Read \& Gilmore}{2005}]{Read_2005}
Read J.~I.,  Gilmore G.,  2005, \mn@doi [MNRAS] {10.1111/j.1365-2966.2004.08424.x}, 356, 107–124

\bibitem[\protect\citeauthoryear{Read, Agertz  \& Collins}{Read et~al.}{2016}]{Read_2016}
Read J.~I.,  Agertz O.,   Collins M. L.~M.,  2016, \mn@doi [MNRAS] {10.1093/mnras/stw713}, 459, 2573–2590

\bibitem[\protect\citeauthoryear{Read, Iorio, Agertz  \& Fraternali}{Read et~al.}{2017}]{Read_2017}
Read J.~I.,  Iorio G.,  Agertz O.,   Fraternali F.,  2017, \mn@doi [MNRAS] {10.1093/mnras/stx147}, p. stx147

\bibitem[\protect\citeauthoryear{Read, Walker  \& Steger}{Read et~al.}{2019}]{Read_2019}
Read J.~I.,  Walker M.~G.,   Steger P.,  2019, \mn@doi [MNRAS] {10.1093/mnras/sty3404}, 484, 1401–1420

\bibitem[\protect\citeauthoryear{{Rey}, {Pontzen}, {Agertz}, {Orkney}, {Read}, {Saintonge}  \& {Pedersen}}{{Rey} et~al.}{2019}]{Rey19}
{Rey} M.~P.,  {Pontzen} A.,  {Agertz} O.,  {Orkney} M. D.~A.,  {Read} J.~I.,  {Saintonge} A.,   {Pedersen} C.,  2019, \mn@doi [\apjl] {10.3847/2041-8213/ab53dd}, \href {https://ui.adsabs.harvard.edu/abs/2019ApJ...886L...3R} {886, L3}

\bibitem[\protect\citeauthoryear{Rey, Pontzen, Agertz, Orkney, Read  \& Rosdahl}{Rey et~al.}{2020}]{Rey2020}
Rey M.~P.,  Pontzen A.,  Agertz O.,  Orkney M. D.~A.,  Read J.~I.,   Rosdahl J.,  2020, \mn@doi [MNRAS] {10.1093/mnras/staa1640}, 497, 1508

\bibitem[\protect\citeauthoryear{Rey, Pontzen, Agertz, Orkney, Read, Saintonge, Kim  \& Das}{Rey et~al.}{2022}]{Rey_2022}
Rey M.~P.,  Pontzen A.,  Agertz O.,  Orkney M. D.~A.,  Read J.~I.,  Saintonge A.,  Kim S.~Y.,   Das P.,  2022, \mn@doi [MNRAS] {10.1093/mnras/stac502}, 511, 5672–5681

\bibitem[\protect\citeauthoryear{Rocha, Peter, Bullock, Kaplinghat, Garrison-Kimmel, Oñorbe  \& Moustakas}{Rocha et~al.}{2013}]{Rocha_2013}
Rocha M.,  Peter A. H.~G.,  Bullock J.~S.,  Kaplinghat M.,  Garrison-Kimmel S.,  Oñorbe J.,   Moustakas L.~A.,  2013, \mn@doi [MNRAS] {10.1093/mnras/sts514}, 430, 81–104

\bibitem[\protect\citeauthoryear{Roper, Oman, Frenk, Benítez-Llambay, Navarro  \& Santos-Santos}{Roper et~al.}{2023}]{Roper_2023}
Roper F.~A.,  Oman K.~A.,  Frenk C.~S.,  Benítez-Llambay A.,  Navarro J.~F.,   Santos-Santos I. M.~E.,  2023, \mn@doi [MNRAS] {10.1093/mnras/stad549}, 521, 1316–1336

\bibitem[\protect\citeauthoryear{Rosdahl, Blaizot, Aubert, Stranex  \& Teyssier}{Rosdahl et~al.}{2013}]{Rosdahl2013}
Rosdahl J.,  Blaizot J.,  Aubert D.,  Stranex T.,   Teyssier R.,  2013, \mn@doi [MNRAS] {10.1093/mnras/stt1722}, 436, 2188

\bibitem[\protect\citeauthoryear{{Ruiz-Lara} et~al.,}{{Ruiz-Lara} et~al.}{2021}]{Ruiz-Lara_2021}
{Ruiz-Lara} T.,  et~al., 2021, \mn@doi [\mnras] {10.1093/mnras/staa3871}, \href {https://ui.adsabs.harvard.edu/abs/2021MNRAS.501.3962R} {501, 3962}

\bibitem[\protect\citeauthoryear{Schive, Liao, Woo, Wong, Chiueh, Broadhurst  \& Hwang}{Schive et~al.}{2014}]{Schive_2014}
Schive H.-Y.,  Liao M.-H.,  Woo T.-P.,  Wong S.-K.,  Chiueh T.,  Broadhurst T.,   Hwang W.-Y.~P.,  2014, \mn@doi [Physical Review Letters] {10.1103/physrevlett.113.261302}, 113

\bibitem[\protect\citeauthoryear{{Schneider}, {Smith}, {Macci{\`o}}  \& {Moore}}{{Schneider} et~al.}{2012}]{Schneider12}
{Schneider} A.,  {Smith} R.~E.,  {Macci{\`o}} A.~V.,   {Moore} B.,  2012, \mn@doi [\mnras] {10.1111/j.1365-2966.2012.21252.x}, \href {https://ui.adsabs.harvard.edu/abs/2012MNRAS.424..684S} {424, 684}

\bibitem[\protect\citeauthoryear{Shipp et~al.,}{Shipp et~al.}{2018}]{Shipp_2018}
Shipp N.,  et~al., 2018, \mn@doi [The Astrophysical Journal] {10.3847/1538-4357/aacdab}, 862, 114

\bibitem[\protect\citeauthoryear{Simon}{Simon}{2019}]{Simon_2019}
Simon J.~D.,  2019, \mn@doi [Annual Review of Astronomy and Astrophysics] {10.1146/annurev-astro-091918-104453}, 57, 375–415

\bibitem[\protect\citeauthoryear{Stopyra, Pontzen, Peiris, Roth  \& Rey}{Stopyra et~al.}{2021}]{Stopyra2021}
Stopyra S.,  Pontzen A.,  Peiris H.,  Roth N.,   Rey M.~P.,  2021, \mn@doi [ApJS] {10.3847/1538-4365/abcd94}, 252, 28

\bibitem[\protect\citeauthoryear{Teyssier}{Teyssier}{2002}]{Teyssier2002}
Teyssier R.,  2002, \mn@doi [A\&A] {10.1051/0004-6361:20011817}, 385, 337

\bibitem[\protect\citeauthoryear{Teyssier, Pontzen, Dubois  \& Read}{Teyssier et~al.}{2013}]{Teyssier_2013}
Teyssier R.,  Pontzen A.,  Dubois Y.,   Read J.~I.,  2013, \mn@doi [MNRAS] {10.1093/mnras/sts563}, 429, 3068–3078

\bibitem[\protect\citeauthoryear{{Tollet} et~al.,}{{Tollet} et~al.}{2016a}]{Tollet16}
{Tollet} E.,  et~al., 2016a, \mn@doi [\mnras] {10.1093/mnras/stv2856}, \href {https://ui.adsabs.harvard.edu/abs/2016MNRAS.456.3542T} {456, 3542}

\bibitem[\protect\citeauthoryear{Tollet et~al.,}{Tollet et~al.}{2016b}]{Tollet_2016}
Tollet E.,  et~al., 2016b, \mn@doi [MNRAS] {10.1093/mnras/stv2856}, 456, 3542–3552

\bibitem[\protect\citeauthoryear{Tulin \& Yu}{Tulin \& Yu}{2018}]{Tulin_2018}
Tulin S.,  Yu H.-B.,  2018, \mn@doi [Physics Reports] {10.1016/j.physrep.2017.11.004}, 730, 1–57

\bibitem[\protect\citeauthoryear{Tweed, Devriendt, Blaizot, Colombi  \& Slyz}{Tweed et~al.}{2009}]{Tweed2009}
Tweed D.,  Devriendt J.,  Blaizot J.,  Colombi S.,   Slyz A.,  2009, \mn@doi [A\&A] {10.1051/0004-6361/200911787}, 506, 647

\bibitem[\protect\citeauthoryear{Weerasooriya, Bovill, Benson, Musick  \& Ricotti}{Weerasooriya et~al.}{2023}]{Weerasooriya_2023}
Weerasooriya S.,  Bovill M.~S.,  Benson A.,  Musick A.~M.,   Ricotti M.,  2023, \mn@doi [The Astrophysical Journal] {10.3847/1538-4357/acc32b}, 948, 87

\bibitem[\protect\citeauthoryear{Wheeler et~al.,}{Wheeler et~al.}{2019}]{Wheeler_2019}
Wheeler C.,  et~al., 2019, \mn@doi [MNRAS] {10.1093/mnras/stz2887}, 490, 4447–4463

\bibitem[\protect\citeauthoryear{Yoshida, Springel, White  \& Tormen}{Yoshida et~al.}{2000}]{Yoshida_2000}
Yoshida N.,  Springel V.,  White S. D.~M.,   Tormen G.,  2000, \mn@doi [The Astrophysical Journal] {10.1086/317306}, 544, L87–L90

\bibitem[\protect\citeauthoryear{Zoutendijk, Brinchmann, Bouché, den Brok, Krajnovic, Kuijken, Maseda  \& Schaye}{Zoutendijk et~al.}{2021}]{Zoutendijk_2021}
Zoutendijk S.,  Brinchmann J.,  Bouché N.,  den Brok M.,  Krajnovic D.,  Kuijken K.,  Maseda M.,   Schaye J.,  2021, \mn@doi [A&A] {10.1051/0004-6361/202040239}

\makeatother
\end{thebibliography}



\appendix

\section{Contrasting \textsc{edge}1 and \textsc{edge}2 star formation histories} \label{appendixA}

As discussed in Section \ref{sec:simulations}, several changes in the sub-grid physics have been introduced between the \textsc{edge}1 suite and the new \textsc{edge}2 iterations. Here we look at how the response that changing the feedback recipe has on the star formation histories.

\begin{figure}
     \centering
      \includegraphics[width=\columnwidth]
      {{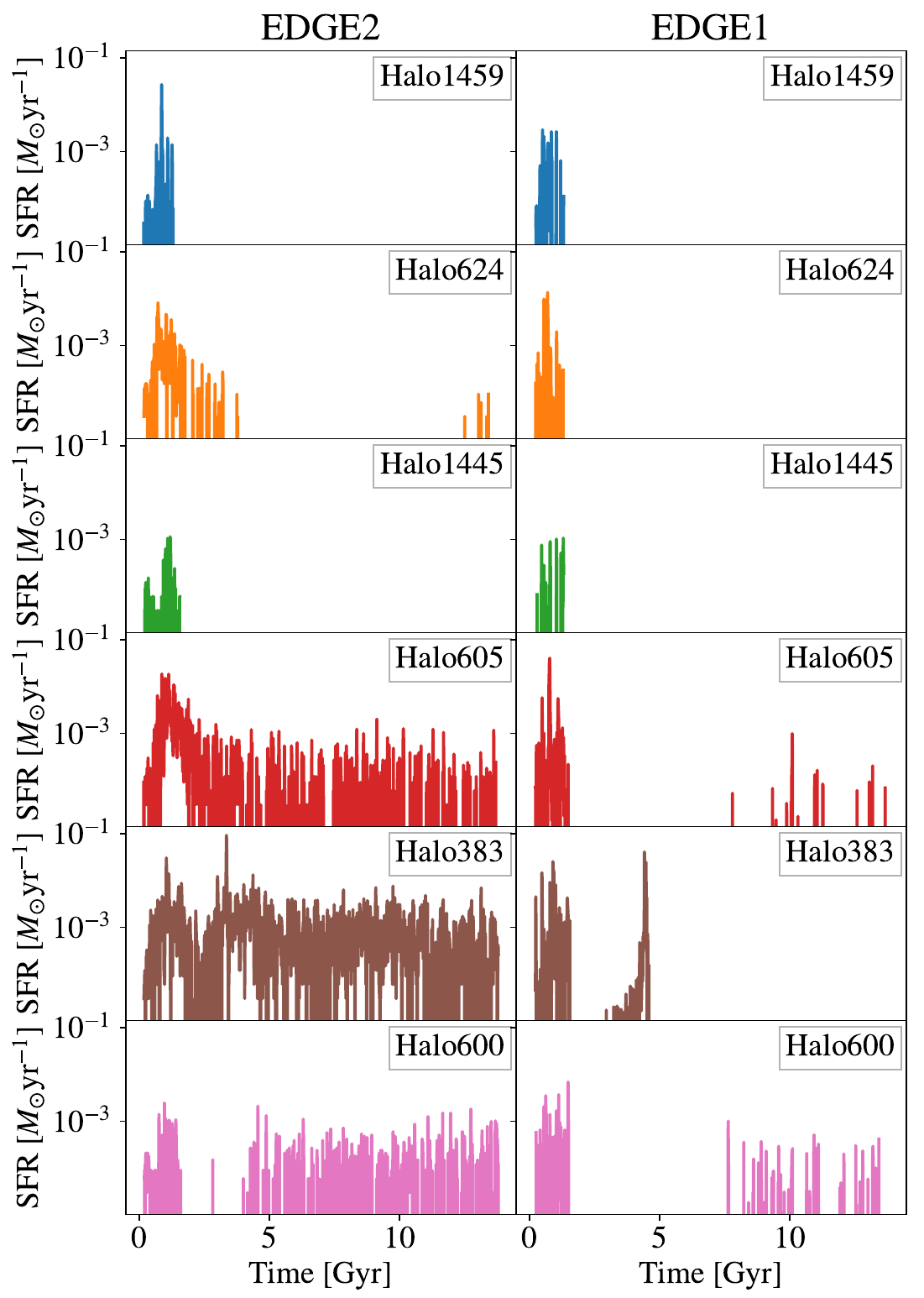}}
     \caption{Comparison between the star formation histories of haloes in \textsc{edge}1 and in \textsc{edge}2. The addition of radiative transfer in \textsc{edge}2 weakens early large-scale outflows and keeps the gas at lower temperatures, allowing more extended star formation until $z=0$.}
     \label{figure_comparison_edge1_SFHs}
\end{figure}

Six of the eight haloes implemented in \textsc{edge}2 have equivalents in the \textsc{edge}1 runs. Fig.~\ref{figure_comparison_edge1_SFHs} shows a direct comparison between the star formation histories of the haloes in the two simulations suites. The \textsc{edge}1 runs see intense large-scale outflows at high redshift which are driven by supernovae, leading to a very hot circumgalactic medium. The addition of radiative transfer in \textsc{edge}2 weakens these outflows and keeps the gas at lower temperatures which, especially combined with the addition of non-equilibrium cooling physics, allows more stars to form. These changes have a clear effect on the SFH, particularly sustaining star formation at late times. Although this is less obvious in the cuspier haloes (1459, 624, 1445), the ones with the strongest cores in \textsc{edge}2 (383, 600, 605) have much greater post-reionisation star formation activity than in their \textsc{edge}1 counterparts. As discussed in the main text, this difference leads directly to \textsc{edge}1 halos remaining cuspier than their \textsc{edge}2 counterparts, in a way that cannot be captured by $\textrm{M}_{\star}/\textrm{M}_{200}$.

\bsp	
\label{lastpage}
\end{document}